\documentclass[amsmath, amssymb, aps, prb, superscriptaddress, longbibliography, preprint]{revtex4-2} 

\usepackage[hmargin=2cm,vmargin=2.5cm]{geometry}
\usepackage[pdftex]{color}
\usepackage{verbatim}
\pdfoutput=1

\usepackage{booktabs,microtype,afterpage} 

\usepackage{graphics}
\usepackage{amssymb}
\usepackage{bm}
\usepackage[charter,greekuppercase=italicized]{mathdesign}

\usepackage[colorlinks=true, allcolors=blue]{hyperref}

\usepackage[charter,greekuppercase=italicized]{mathdesign}
\definecolor{red}{rgb}{0.85,.1,0}
\definecolor{green}{rgb}{0.0,0.6,0.0}
\definecolor{orange}{rgb}{1,0.5,0}
\usepackage{graphicx}    
\graphicspath{{./}{figure/}}
\DeclareGraphicsExtensions{.eps,.png,.pdf,.jpg}

\begin{document}

\title{Resistive Switching and Neuromorphic Computing in Metal/Nb:SrTiO$_3$: Mechanisms, Interface Physics, and Charge Transport}

\author{Christopher Broyles}
\thanks{Corresponding author: cbroyles@lanl.gov}
\affiliation{Center for Integrated Nanotechnologies, Los Alamos National Laboratory, Los Alamos, NM, 87545, USA}

\author{Elizabeth Krenkel}

\affiliation{Center for Integrated Nanotechnologies, Los Alamos National Laboratory, Los Alamos, NM, 87545, USA}

\author{Frank Barrows}
\affiliation{Theoretical Division T-4, Los Alamos National Laboratory, Los Alamos, NM, 87545, USA}

\author{Sundar Kunwar}
\affiliation{Center for Integrated Nanotechnologies, Los Alamos National Laboratory, Los Alamos, NM, 87545, USA}

\author{Aiping Chen}
\thanks{Corresponding author: apchen@lanl.gov}
\affiliation{Center for Integrated Nanotechnologies, Los Alamos National Laboratory, Los Alamos, NM, 87545, USA}

\date{\today}
\begin{abstract}


Resistive switching (RS) in Nb-doped SrTiO$_3$ (Nb:STO) based memristive devices has attracted sustained interest in information processing and novel computing because of its forming-free operation, large on/off ratio, and gradual conductance modulation. Metal/Nb:STO Schottky junctions have emerged as a prototypical system for understanding RS mechanisms. Despite more than two decades of research, the physical origin of RS remains controversial, with proposed mechanisms including charge trapping and detrapping, oxygen vacancy migration, tunneling, interfacial redox reactions, and conductive filament formation. In this review, we examine these seemingly competing mechanisms and show that many experimental observations can be understood within a unified framework centered on the formation and evolution of an extrinsic interfacial layer at the metal/Nb:STO interface. We discuss how interface quality, interface inhomogeneity and defect-mediated processes, including proton incorporation, oxygen vacancy dynamics, and tunneling, govern Schottky barrier modulation and RS behavior. We further summarize how fabrication conditions, measurement protocols, and aging influence the interface formation and switching characteristics. This review establishes an integrated picture of M/Nb:STO heterojunctions and provides design principles for reliable oxide memristive devices through interface and defect engineering in M/Nb:STO and M/oxide/Nb:STO systems. 

\end{abstract}

\maketitle

\clearpage

\section{Introduction}

\begin{figure}[t]
    \centering
    \includegraphics[width=0.5\linewidth]{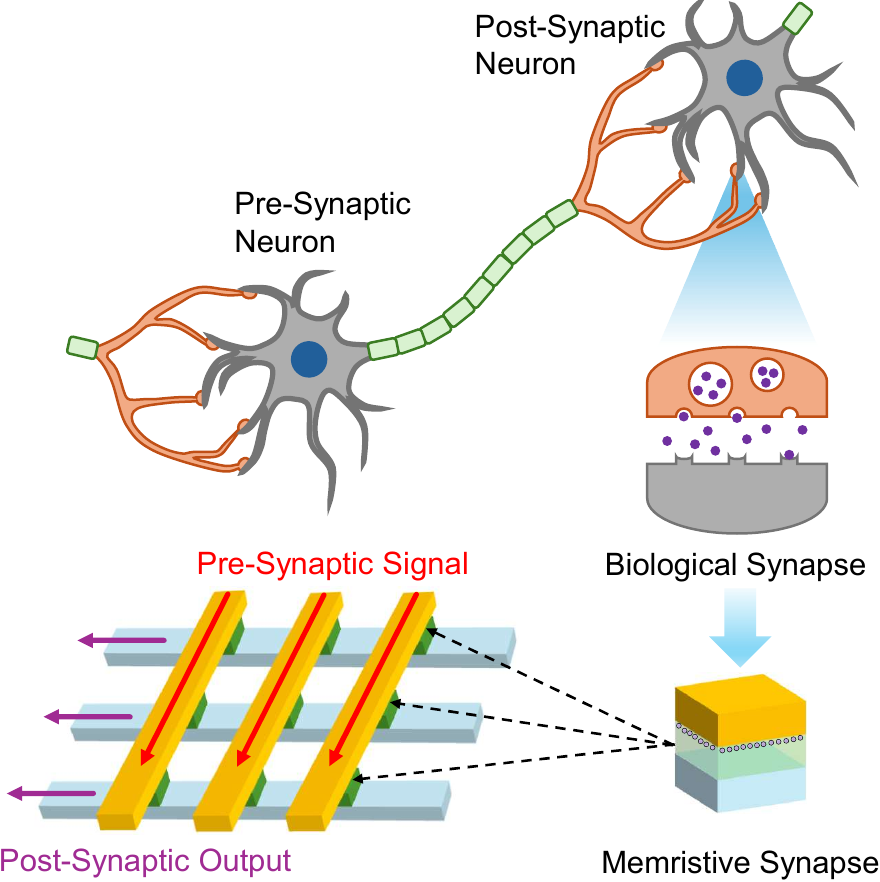}
    \caption{Comparative diagram of biological neuron and synapse relating to memristive synapse integrated in crossbar array.}
    \label{fig:1}
\end{figure}

\subsection{Oxides for Resistive Memory and Neuromorphic Systems}

Resistive switching (RS) refers to the electrically induced, reversible transition between distinct resistance states in a device, with the resistance determined by the history of the applied electrical stimulus. RS phenomena have been extensively studied over the past few decades due to their promising applications in information storage, computing, and other emerging information technologies. Metal oxides, particularly perovskites such as SrTiO$_3$ (STO), have emerged as promising materials for RS devices.~\cite{Sawa2008MaterialsToday}  This electrically controlled modulation of resistance forms the basis of resistive random-access memory (ReRAM), an emerging nonvolatile memory technology that offers several attractive features, including high scalability, low power consumption, and fast switching speeds.~\cite{Ielmini2025ChemicalReviews,Slesazeck2019Nanotechnology,Li2018JoPDAP,Jeong2012RPP} 
Unlike conventional flash memory, which relies on charge storage and requires high programming voltages, ReRAM operates through alternative mechanisms that enable faster, low-voltage switching and improved endurance.~\cite{Slesazeck2019Nanotechnology} These characteristics make ReRAM not only a potential replacement for flash but also a promising platform for in-memory computing in post–von Neumann architectures.~\cite{Seok2024AdvancedElectronicMaterials} RS devices have been widely proposed for neuromorphic computing, as they emulate the adaptive behavior of biological synapses and neurons.~\cite{Li2018JoPDAP,Rao2023N, Xiao2025AFM,Mohanan2024Nanomaterials}

Different from the von Neumann architecture, neuromorphic computing systems are inspired by the brain and made of interconnected artificial synapses and neurons (Fig.~\ref{fig:1}). The neurons, based on their nonlinear dynamics, generate spikes (action potentials) that provide the main data representation and communication mechanism.~\cite{Vinck2023N} Computational tasks are distributed across the neural network where synapses implement both the memory and the computational units, enabling in-memory computing.~\cite{Hassabis2017N,Mehonic2020AIS,Burr2017APX} These synapses obtain their weights by means of learning mechanisms such as spike-timing-dependent plasticity (STDP), which is a local learning algorithm first discovered in biological synapses.~\cite{Burr2017APX,Saighi2015FrontiersInNeuroscience} Fig.~\ref{fig:1} shows neurons connected by many synapses in biological systems, and the analogous memristive crossbar arrays, which can be used to mimic the biological synapses.~\cite{Xia2019NM, Zhang2018PSSA, Li2021AIS} Since the experimental demonstration of TiO$_2$ memristors,~\cite{Strukov2008Nature, Yang2008NN} this field has attracted significant attention in the past two decades.

\subsection{Current vs. Voltage Rotation Sequence and Switching Mechanisms}
\label{sec:1B}

\begin{figure}[t]
    \centering
    \includegraphics[width=1\linewidth]{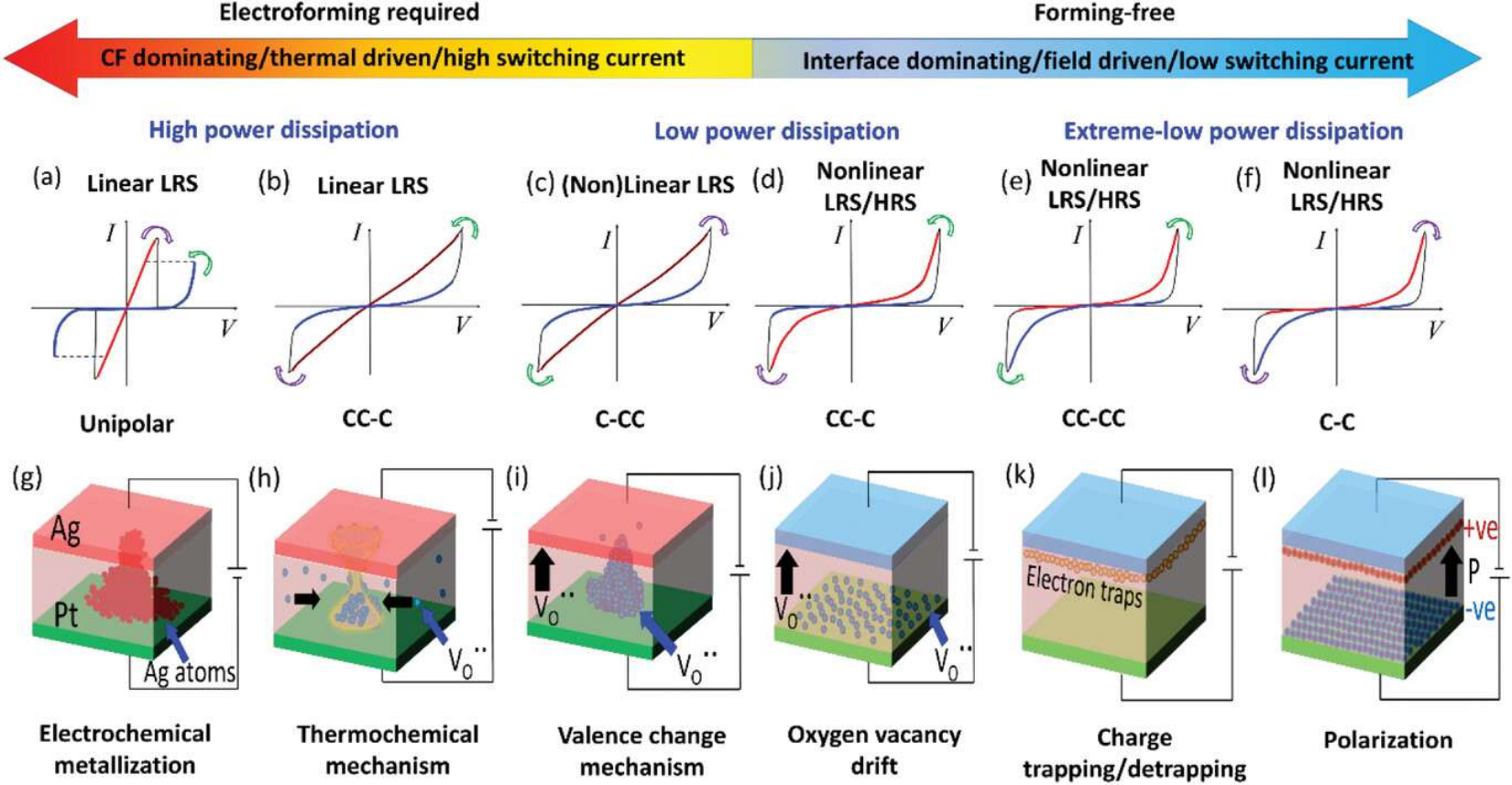}
    \caption{Schematic illustration of different types of resistive switching mechanisms observed in memristors using oxide materials as RS layers. From left to right shows filament-type and interface-type switching with different types of I-V characteristics along with the possible switching mechanisms. Figure taken from Ref.~\cite{Roy2022AEM}.}
    \label{fig:2}
\end{figure}

In general, RS devices can be divided into filament-type switching and interface-type switching, where RS occurs between a high resistance state (HRS) and low resistance state (LRS). Figure \ref{fig:2} summarizes the basic current vs. voltage (I-V) characteristics for different types of devices with possible mechanisms. In general, filament-type devices require an electroforming process and possess high power dissipation as it is dominated by conducting filament (CF).~\cite{Jeong2012RPP, Zhang2023AMI, Zhang2022ComputationalMaterials} The migration of defects (e.g., metal ions and/or oxygen vacancies) plays a pivotal role in the formation and dynamics of CFs.~\cite{Perez2024AdvancedElectronicMaterials} Filament-type switching, including thermochemical mechanism (TCM, Fig.~\ref{fig:2}a and h), electrochemical metallization (ECM, Fig.~\ref{fig:2}b and g),  and valence change mechanism (VCM, Fig.~\ref{fig:2}b, c, d and i), have been widely discussed and summarized in many comprehensive reviews.~\cite{Sawa2008MaterialsToday, Waser2009AM, Dittmann2021AP, Ielmini2025ChemicalReviews,  Slesazeck2019Nanotechnology, Seok2024AdvancedElectronicMaterials,Li2020JoMCC,Mohammad2016NanotechnologyReviews, Kim2025AMI} 
To accommodate more recent developments in interface-type switching with different I-V rotation sequences, we use the current rotation direction at first quadrant and the third quadrant in linear I-V plots as an alternative way to define I-V rotation sequence. The figure-eight is equivalent to counterclockwise-clockwise (CC-C) and counter-figure-eight corresponds to clockwise-counterclockwise (C-CC). 

All above mentioned filament-type memristors often require an electroforming process. Interface-type memristors, on the other hand, are often forming-free~\cite{Sawa2008MaterialsToday, Bagdzevicius2022, Kunwar2023AIM}. They may offer the advantage of superior stability, non-linear current with self-rectification characteristics, low power dissipation, and area scalability. ~\cite{Sawa2008MaterialsToday, Bagdzevicius2022, Kunwar2023AIM} Interface-type switching is often related to the physical or chemical phenomena occurring at the oxide/electrode interface.~\cite{Bagdzevicius2022} The most observable feature of such a switching process is the scaling of the device resistance with the electrode area. ~\cite{Baeumer2016N, Gutsche2021FrontiersinNeuroscience, Antola2025AppliedElectronicMaterials} In the case of interface-type RS devices, three switching mechanisms are often considered, as shown on the right side of Fig.~\ref{fig:2}: (i) OV migration under the whole electrode (interface-type VCM)~(see Fig.~\ref{fig:2}j);~\cite{Zhang2026AFM} (ii) electronic and/or electrostatic effect~(see Fig.~\ref{fig:2}k);  and (iii) polarization-induced barrier modification~(see Fig.~\ref{fig:2}l).~\cite{Bagdzevicius2022, Chen2020AFM} The RS in M/Nb:STO has often related to electronic and electrostatic mechanisms (e.g., charge trapping/detrapping)~\cite{Mikheev2014NatureCommunications, Fan2017JoMCC} and filament-type VCM.\cite{Baeumer2016N}  In both cases, I-V hysteresis shows CC-C rotation sequence. The dynamics of space-charges arising from trapping/detrapping of carriers can modulate the Schottky barrier height (SBH) and depletion width ($W_\mathrm{d}$), resulting  in the characteristic CC-C I-V hysteresis observed in (Au, Pt, Ni)/Nb:STO.~\cite{Mikheev2014NatureCommunications,Bourim2014JSSST, Fan2017JoMCC,Shen2013APA, Yin2015PCCP, Goossens2018JAP}

It is interesting that these mechanisms show a variety of I-V current rotation sequences as defined in Fig.~\ref{fig:2}a-f. Some of them are relatively well understood, such as ECM and TCM, whereas others are more difficult to interpret. For example, both CC-C and C-CC I-V rotations have been reported for VCM. Dittmann and co-workers reconciled these seemingly opposite switching polarities by considering the competition between oxygen-vacancy redistribution within the oxide and oxygen exchange at the active electrode.~\cite{Dittmann2021AP}  When vacancy concentration polarization near the active interface dominates, the device typically exhibits CC-C switching; when interfacial oxygen exchange becomes dominant, the switching polarity can reverse, producing C-CC behavior. Thus, the I-V rotation in VCM is not determined simply by the direction of oxygen-vacancy migration, but rather by how the resulting defect redistribution and interfacial redox reactions modify the dominant electronic barrier. Despite these efforts, establishing a general framework that connects different switching mechanisms to their characteristic I-V rotation sequences remains an important direction for future research. Such a framework should systematically consider the interface type (Schottky or Ohmic), number of active interfaces (single or double Schottky), carrier type (n- or p-type), physical origin of switching (electronic, ionic, or coupled electronic-ionic), spatial nature of switching (filamentary, interfacial, or mixed), and the spatial distribution and dynamics of defects.

\subsection{Why Metal/Nb:STO}

\begin{figure}
    \centering
    \includegraphics[width=0.5\linewidth]{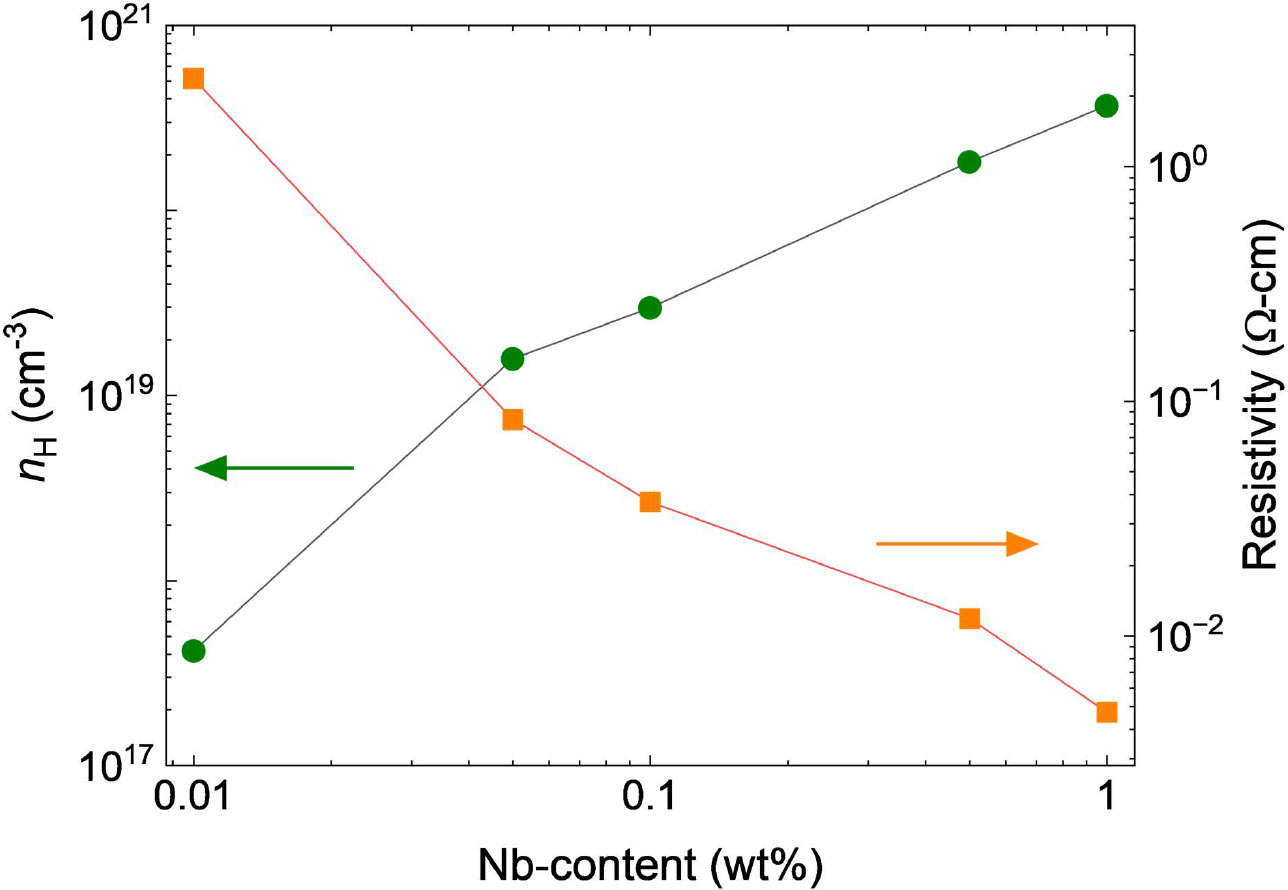}
    \caption{The effect of Nb-doping on carrier concentration calculated from Hall measurements ($n_\mathrm{H}$) and room temperature resistivity. The figure has been digitized from Ref.~\cite{Fujii2007PRB}.}
    \label{fig:3}
\end{figure}

STO is a widely used substrate for epitaxial growth of perovskite oxide films. When doped with niobium (Nb), STO becomes an n-type semiconductor. As shown in Fig.~\ref{fig:3}, lightly doped crystals (i.e., 0.05–0.1 wt\% Nb) typically exhibit resistivity on the order of 0.1–1 $\mathrm{\Omega} \cdot$cm, while higher doping levels (i.e., 0.5–1.0 wt\%) reduce the resistivity to the order of 0.01 $\mathrm{\Omega} \cdot$cm. This low resistivity has enabled it to serve as conducting bottom electrodes for epitaxial growth of many perovskite oxide thin films with good lattice matching. The carrier concentration is  10$^{18}$ – 10$^{21}$ cm$^{-3}$ depending on doping (Fig.~\ref{fig:3}), while a metal generally has electron densities on the order of 10$^{22}$ cm$^{-3}$. 
Nb-doped STO (Nb:STO) is considered as a nearly degenerate semiconductor. When interfaced with a high-work-function metal, a rectifying Schottky junction is often reported.~\cite{Park2008JAP} Studying such an interface between Nb:STO and metal electrodes or conducting oxide layer is critical to understand the physical properties of heterostructures grown on Nb:STO. In the 90s, some initial work has already reported an interfacial layer with a lower dielectric constant than STO formed at the Au/Nb:STO interface.~\cite{Hasegawa1991JAP, Yoshida1991PhysicaB, Yoshida1991JournalofAppliedPhysics, Shimizu1999JAP} It is very likely such a low permittivity interfacial layer is the origin of the so-called “dead layer” near the ferroelectric surface.~\cite{Zhou1997JournalofAppliedPhysics} The RS effects of the interfacial and depletion layers will be defined and discussed in Sec.~\ref{sec:II}.  

Studying metal contact with semiconducting oxides is one of the key areas of oxide electronics. Noble metals like Pt and Au are widely used for electrical properties characterization of semiconducting, dielectric, and ferroelectric oxide thin films. M/oxide/Nb:STO (M = metal) is a popular geometry and a variety of oxides have been investigated in this format. For example, ferroelectrics such as BaTiO$_3$, BiFeO$_3$, Pb(Zr$_{x}$Ti$_{1-x}$)O$_3$ and their doped variates;~\cite{Wen2014AppliedPhysicsLetters,Yang2008AppliedPhysicsLetters, Bai2016AppliedMaterialsandInterfaces} dielectrics such as HfO$_2$ and LaAlO$_3$;~\cite{Dong2025JournalofVacuumScienceandTechnologyB, Tian2011APA, Gurukrishna2026JMR} and semiconducting oxides such as ZnO and TiO$_2$ have been heavily studied.~\cite{Wu2008APL,Zhu2012JoPDAP,Shivaram2025JMR, Gurukrishna2026JMR} 
However, as stated by Fan \textit{et al}., one of the concerns is that the I-V hysteresis loops of M/oxide/Nb:STO are similar with that in M/Nb:STO systems,~\cite{Fan2017JoMCC} which makes the role of oxide layer sandwiched between M and Nb:STO unclear. Therefore, there is a strong need to look carefully at the basic semiconductor behavior of  M/Nb:STO. Understanding simple  M/Nb:STO systems is a critical step to understand more complex M/oxide/Nb:STO devices and many other M/oxide switching systems.
Although M/Nb:STO heterojunctions are one of the simplest structures for RS, the physical origins of RS are complex and remain controversial.~\cite{Mikheev2014NatureCommunications, Fan2017JoMCC, Li2023APL,Wang2016ASS,Yang2014JAP, Li2010MSEB} The confusion arises from the wide variety of I–V hysteresis loops and switching mechanisms reported even in these seemingly simple systems.~\cite{Buzio2024JoPDAP} Since the first report of M/Nb:STO Schottky barriers in the late 1960s,~\cite{Sroubek1969SSC} more than half a century of sustained research has yielded substantial advances and valuable physical insights into these heterojunctions. However, a comprehensive review that reconciles the proposed switching mechanisms and systematically connects them with fabrication conditions, interface quality, and measurement protocols is still lacking. In this review, we examine the semiconductor physics governing M/Nb:STO interfaces and critically assess the possible origins of their resistive-switching behavior by considering interface physics. This represents a critical step toward establishing a general framework for understanding RS across diverse metal/oxide systems, which is essential for advancing RS devices for emerging computing applications.

\section{Reconciliation of Switching Mechanisms}
\label{sec:II}

\begin{figure}[t]
    \centering
    \includegraphics[width=0.8\linewidth]{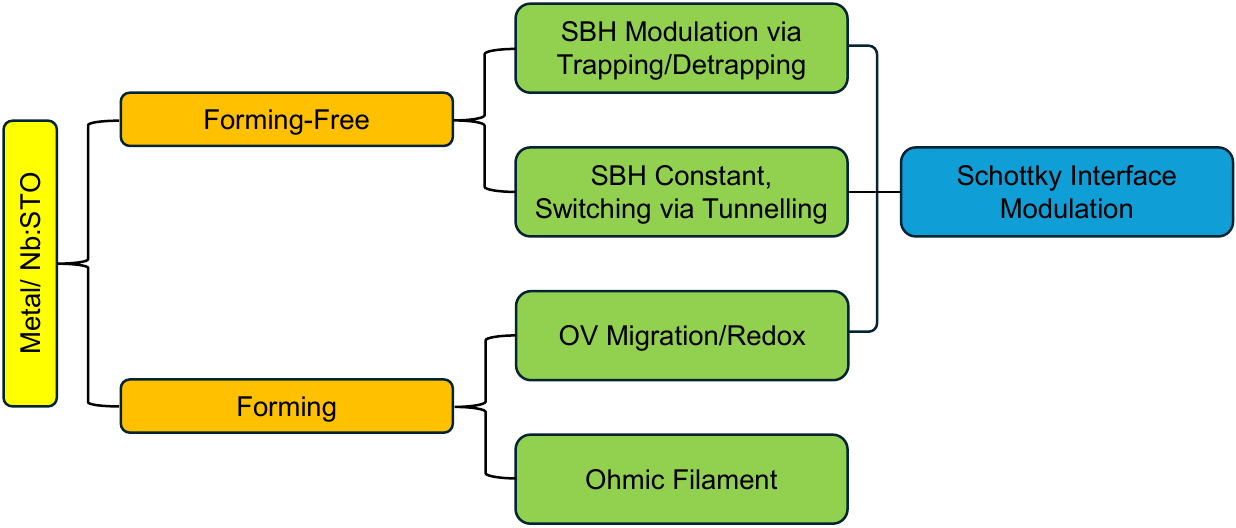}
    \caption{Graphical summary of RS mechanisms in M/Nb:STO.}
    \label{fig:4}
\end{figure}

Decades of research in M/Nb:STO (M = Au, Pt, \textit{etc}.) have resulted in a variety of I-V hysteresis loops and different mechanisms have been proposed. Fig.~\ref{fig:4} summarizes some of most popular mechanisms proposed in literature. Within the filament mechanism, the claim of ohmic filament formation (linear I-V in LRS) in such devices is rare.~\cite{Bian2024JVSTB}  With support of solid experimental evidence, Yang \textit{et al.} and Baeumer \textit{et al}. proposed that valence change mechanism is responsible for the RS after a forming process.~\cite{Baeumer2016N, Yang2014JAP} A handful of papers supported such a redox-based filamentary switching in M/Nb:STO,~\cite{Baeumer2016N, Yang2014JAP,Chen2012JAP, Li2023APL, Wang2013APL} but a lot of other reports pointed out the importance of Schottky interface to enable the switching. Schottky barrier height modulation by charge trapping/detrapping has been widely discussed.~\cite{Brillson2011JAP,Dharanya2022JNP,Bourim2013CAP,Buzio2012APL,Bian2025FML,Bourim2014JSSST,Chen2011APL,Park2014APL,Lee2014APLM,Quinonez2025APLED,Shen2013APA,Zhong2013CAP,Chen2010APL,Mikheev2014NatureCommunications,Buzio2024JoPDAP,Fan2017JoMCC,Park2008JAP,Li2018PSSA,Goossens2018JAP,Kunwar2023AEM,Yin2015PCCP,Kan2013APL,Li2019PSSA} However, some studies have found that the average SBH remains nearly constant and have instead attributed RS to tunneling through spatially inhomogeneous regions of the Schottky barrier.~\cite{Lee2011APL, Fujii2007PRB, Shang2008APL,Shang2009APL, Wang2016ASS, Lee2014APLM}

\begin{figure}[t]
    \centering
    \includegraphics[width=0.7\linewidth]{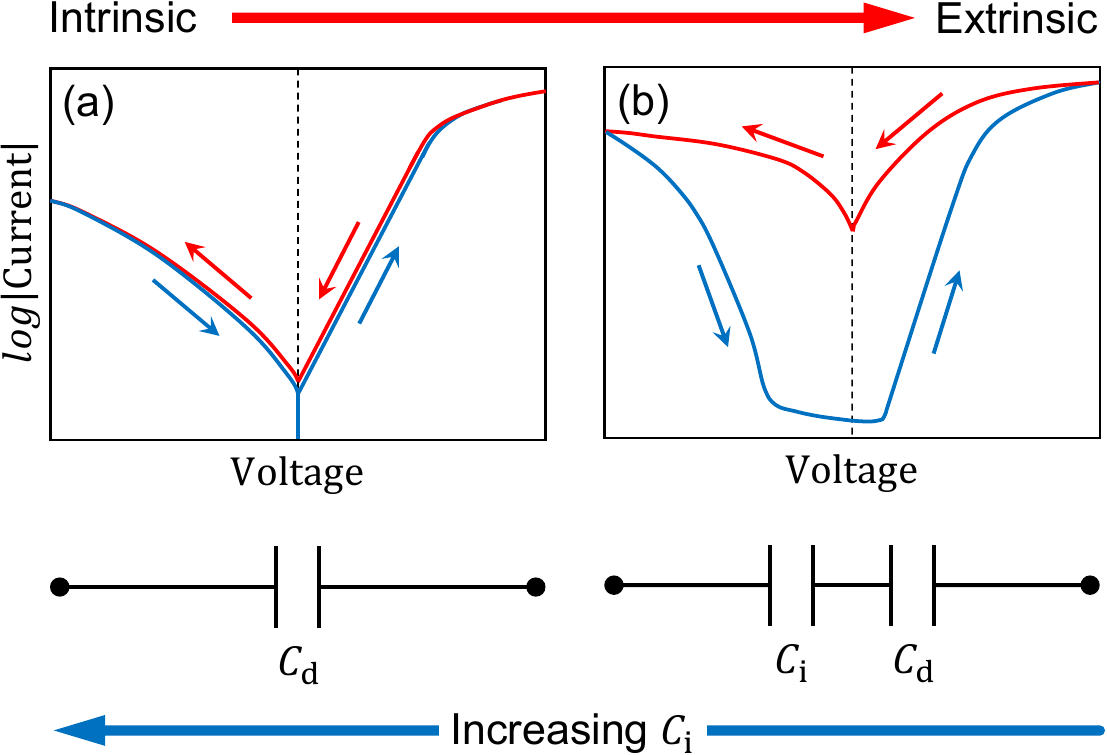}
    \caption{Illustrations of M/Nb:STO I-V curve and equivalent circuits for (a) an intrinsic Schottky junction with a depletion layer capacitance, $\mathrm{C_d}$, in addition to an (b) extrinsic Schottky junction with series interfacial, $C_\mathrm{i}$, and depletion layer capacitance. For all illustrations, positive voltage is forward bias, while negative voltage is reverse bias.}
    \label{fig:5}
\end{figure}

M/Nb:STO is one of a few systems that can show RS without electroforming. The Schottky nature of such a system has been widely reported. Therefore, RS in M/Nb:STO has often been recognized as an interface-type RS switching.~\cite{Brillson2011JAP,Dharanya2022JNP,Bourim2013CAP,Buzio2012APL,Bian2025FML,Bourim2014JSSST,Wang2016ASS,Chen2011APL,Park2014APL,Lee2014APLM,Quinonez2025APLED,Shen2013APA,Zhong2013CAP,Chen2010APL,Mikheev2014NatureCommunications,Buzio2024JoPDAP,Fan2017JoMCC,Park2008JAP,Li2018PSSA,Goossens2018JAP,Kunwar2023AEM,Yin2015PCCP,Kan2013APL,Li2019PSSA,Lee2011APL,Fujii2007PRB,Sawa2005APL,Shang2008APL,Shang2009APL} Fig.~\ref{fig:5} shows two typical I-V hysteresis loops for M/Nb:STO. The I-V rotation sequence is CC-C type. Various works have demonstrated a correlation between the growth method, interface quality and RS behavior, where a high-quality (epitaxial) electrode intimately contacting with annealed Nb:STO leads to the suppression of RS.~\cite{Mikheev2014NatureCommunications, Buzio2024JoPDAP,Li2023APL}. We defined this type of M/Nb:STO interface as the intrinsic limit (Fig.~\ref{fig:5}a). 
Typically, M/Nb:STO Schottky junctions consist of an interfacial layer capacitance ($C_\mathrm{i}$) in series with the depletion layer capacitance ($C_\mathrm{d}$).~\cite{Hasegawa1991JAP, Yoshida1991PhysicaB, Yoshida1991JournalofAppliedPhysics, Shimizu1999JAP} In the intrinsic case, the Schottky interface is dominated by $C_\mathrm{d}$, such that the equivalent circuit can be simplified to $C^{-1} = C_\mathrm{i}^{-1} + C_\mathrm{d}^{-1} \approx C_\mathrm{d}^{-1}$. However, M/Nb:STO interfaces often deviate from above ideal scenario and there is a substantial interfacial layer. We define this type of interface is in the extrinsic limit.  A large barrier modulation results in a large hysteresis loop (Fig.~\ref{fig:5}b), while in the intermediate case, the hysteresis loop is suppressed.~\cite{Mikheev2014NatureCommunications} It is our understanding that the formation of an extrinsic interfacial layer between the metal and Nb:STO is central to the origin of RS in interface-type M/Nb:STO junctions. Within the SBH-modulation picture, electron trapping at and/or within this interfacial layer increases the effective SBH and $W_\mathrm{d}$, driving the junction into the HRS (see blue curve in Fig.~\ref{fig:5}b).  Conversely, electron detrapping under positive (forward) bias reduces the SBH and $W_\mathrm{d}$, leading to the LRS (see red curve in Fig.~\ref{fig:5}b). In this case, the extrinsic $C_\mathrm{i}$ is in series with $C_\mathrm{d}$. Compared with the intrinsic junction limit, the presence of this extrinsic interfacial layer is associated with a smaller $C_\mathrm{i}$, a larger ideality factor, and a larger effective SBH. In the spatially inhomogeneous tunneling picture, by contrast, the SBH remains essentially unchanged, while RS arises from modulation of tunneling through local regions of the interfacial barrier. Although SBH modulation and tunneling may initially appear to be competing mechanisms, we will show later that they can be viewed as two manifestations of the same underlying interfacial physics. Importantly, this extrinsic interface also provides the starting point for filament-type VCM: prior to electroforming, transport in the HRS is still dominated by the extrinsic interfacial barrier. As discussed later, electroforming locally disrupts this Schottky interface and creates a conductive filament, thereby transforming the dominant switching behavior from interface-type RS to filament-type VCM.


\subsection{The Interfacial Layer}
\label{subsec:IIA1}

The temperature ($T$)-dependent barrier heights from I–V and $T$-independent flat-band voltages from C–V were observed in Au/Nb:STO.~\cite{Shimizu1999JAP,Susaki2007PRB} To reconcile such a discrepancy, a model with an interfacial layer at the M/Nb:STO interface was proposed. Hasegawa \textit{et al}. and Yoshida \textit{et al}. independently reported this interfacial layer exhibits a lower permittivity than STO.~\cite{Hasegawa1991JAP,Yoshida1991PhysicaB, Yoshida1991JournalofAppliedPhysics} Later, in a $T$-dependent study, Susaki \textit{et al}. found that this interfacial layer exhibits an opposite $T$ dependence of bulk STO permittivity.~\cite{Susaki2007PRB} Owing to the “quantum paraelectric” nature of STO, its permittivity monotonically increases with decreasing $T$, where $\epsilon_\mathrm{r}(300~K) \approx$~300 and $\epsilon_\mathrm{r}(4~K) \approx$~20,000.~\cite{Muller1979PRB} Therefore, the leakage of STO based devices is expected to significantly reduce with decreasing $T$ due to both increased dielectric constant and reduced thermal activation with decreasing $T$. However, the permittivity of this interfacial layer (within $\approx$5~nm from the interface) decreases with decreasing $T$ (i.e., below 10 at 10K).~\cite{Susaki2007PRB} This results in significant $W_\mathrm{d}$ narrowing.~\cite{Hirose2015APL,Ohsawa2021JoPCC} Along with the reduced permittivity, Susaki \textit{et al}. observed a surprising higher leakage at low $T$ due to tunneling current dominating the reverse bias.~\cite{Susaki2007PRB} This counterintuitive behavior has also been reported in more recent studies.~\cite{Ohsawa2021JoPCC,Kim2020APL,Goossens2018JAP,Sawa2005APL,Suzuki1997JAP} 

\begin{figure}[t]
    \centering
    \includegraphics[width=\linewidth]{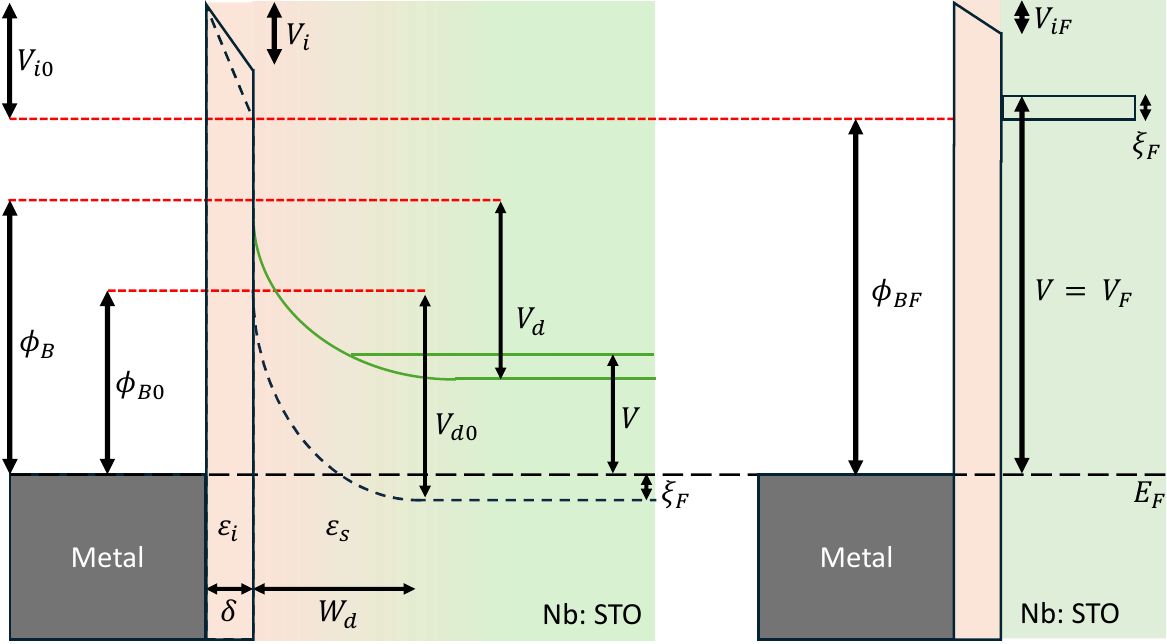}
    \caption{Schottky barrier energy diagram and flat band equivalent schematics, where $V$ indicates a forward bias voltage. The voltage-dependent Schottky barrier height ($\phi_\mathrm{B}$), interfacial voltage ($V_\mathrm{i}$), and depletion voltage ($V_\mathrm{d}$) are denoted in the zero-bias and flat-band limits by the addition of the subscript "$0$" or "$F$", respectively.~\cite{Yamamoto1998JJAP} }
    \label{fig:7}
\end{figure}

The metal-insulator-semiconductor (MIS) interfacial model is motivated by observations of the low-dielectric permittivity interfacial layer and contradicting barrier heights from C-V and I-V measurements, which indicate a voltage-dependent barrier height. For Nb:STO, mechanisms such as a thin insulating layer, surface states, or structural imperfections cause the applied voltage to become shared between the interfacial ($V_\mathrm{i}$) and depletion ($V_\mathrm{d}$) layers, as seen in Fig.~\ref{fig:7}.~\cite{Yamamoto1998JJAP} To overcome a voltage-dependent barrier height, Wagner \textit{et al}.~\cite{Wagner1983IEEE} and Yamamoto \textit{et al}.~\cite{Yamamoto1998JJAP} made the connection to the flat-band barrier height ($\phi_{BF}$) in the limit of zero-electric field across the semiconductor (i.e., no depletion region), given by flat energy bands (see Fig.~\ref{fig:7}). Ultimately, the MIS model provides a electrostatic framework to relate measured parameters from I-V fittings to a voltage-independent $\phi_\mathrm{BF}$.

\begin{figure}[t]
    \centering
    \includegraphics[width=\linewidth]{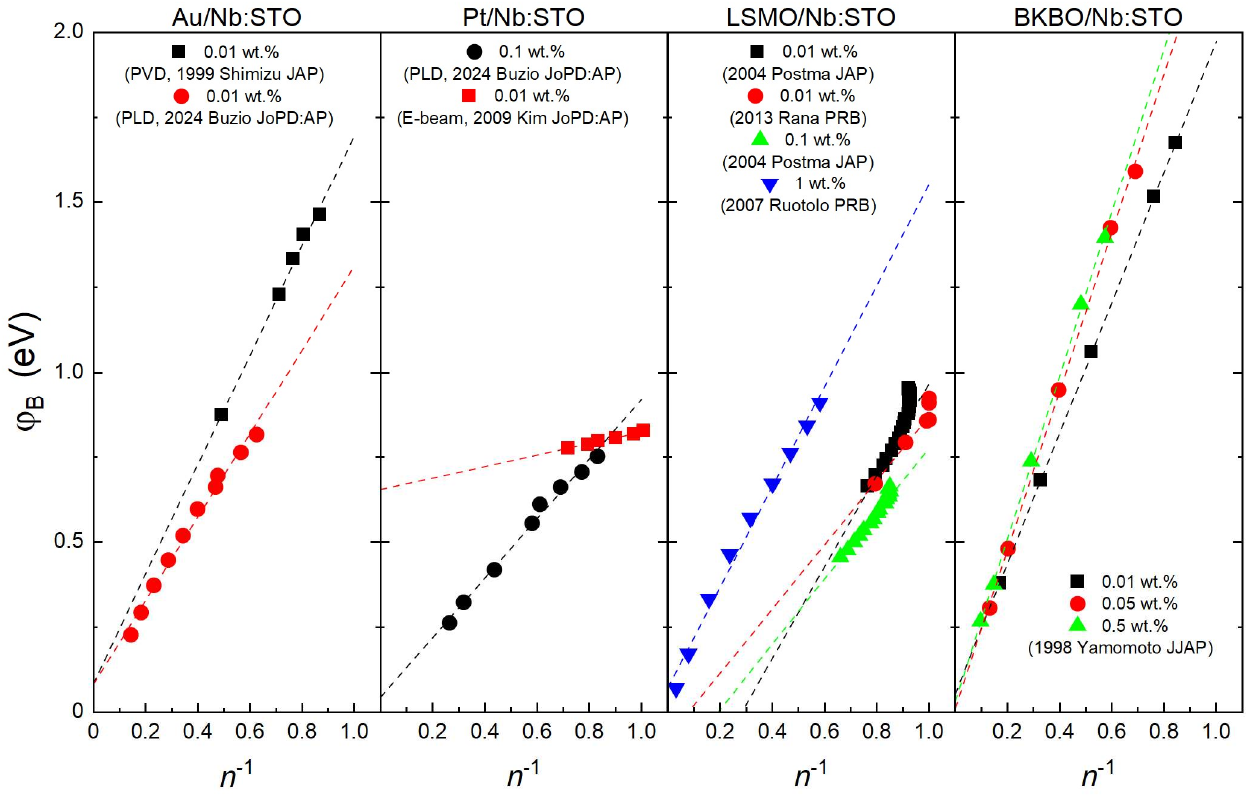}
    \caption{ Linear relationship for the temperature-dependence of $\mathrm{\phi_B}$ vs. $n^{-1}$ for M and conductive oxide electrodes on Nb:STO. The data has been digitized from the respective references (see Tab.~\ref{tab:MIS_fittings}).}
    \label{fig:6}
\end{figure}

Within the MIS transport model, the insulating interfacial layer is supported by the linear relationship between the SBH ($\mathrm{\phi_B}$) and ideality factor ($n$), which can be obtained from thermionic emission fittings of the forward bias I-V measurement~(see Fig.~\ref{fig:6}).~\cite{Yamamoto1998JJAP, Wagner1983IEEE}  The conventional forward bias thermionic emission fitting can be altered because the voltage bias becomes shared between the interfacial and depletion layer (i.e., $V = V_\mathrm{i}+V_\mathrm{d}$), in addition to other effects such as image force lowering.~\cite{Yamamoto1998JJAP} Yamamoto \textit{et al}. quantified this effect on $n$ by assuming constant voltage dependence of the SBH in the following equation:
\begin{equation}
    n^{-1}=1-\frac{d\phi_B}{dV} = 1-\frac{\phi^v_B -\phi_{B}}{V},
    \label{eqn:1}
\end{equation}
where $\phi^v_B$ and $\phi_{B}$ are the voltage-dependent and zero-bias values of the SBH, respectively.~\cite{Yamamoto1998JJAP} Next, Yamamoto \textit{et al}. connected this equation with the flat-band energy diagram of the Schottky barrier (i.e., $V_F = \phi_{BF} + \xi_F$). The flat-band barrier height ($\phi_{BF}$) occurs at the voltage where the depletion width ($W_\mathrm{d}$) equals zero.~\cite{Yamamoto1998JJAP}  Combining the conventional and flat-band frameworks, (Fig.~\ref{fig:7}) the linear fitting can be achieved by solving for $\phi_{B}$:
\begin{equation}
    \phi_{B} = (\phi_{BF} + \xi_F)n^{-1} - \xi_F.
    \label{eqn:2}
\end{equation}
The broad alignment of this interfacial layer fitting across M and conductive contacts on Nb:STO complements the capacitance measurements that indicate the existence of an interfacial layer (see Fig.~\ref{fig:6}).

The fitting parameters of $\phi_\mathrm{BF}$ and $\xi_\mathrm{F}$ are shown in Tab.~\ref{tab:MIS_fittings}, and further information can be gathered from the validity of this fittings. In general, $\phi_\mathrm{BF}$ is expected to equal the built-in potential from C-V measurements and $\xi_\mathrm{F} \approx$ 100 meV for M/Nb:STO.~\cite{Hikita2007APL} A larger $\phi_{\mathrm{BF}}$ obtained from MIS I-V relative to C-V typically indicates enhanced interfacial effects, such as the presence of an extrinsic interfacial layer as opposed to an intrinsic one. Conversely, a smaller MIS-extracted $\phi_{\mathrm{BF}}$ suggests the contribution of tunneling, which increases current through thin or defective interfacial regions and reduces the apparent barrier height. Additionally, unusually large values of $\xi_\mathrm{F}$ point to non-ideal interface conditions, such as defect-induced surface doping, oxygen vacancies, or interfacial layers introduced during electrode deposition. These limits are typical for comparing I-V and C-V measurements of the SBH, since capacitance measurements probe the spatially averaged response instead of spatial variations. 


\begin{table}[t]
\caption{Summary of extracted flat-band Schottky parameters for linear fittings in Fig.\ref{fig:6}.}
\label{tab:MIS_fittings}
\begin{ruledtabular}
\begin{tabular}{lccccc}
Article & Electrode & Nb wt.\% & $\phi_{\mathrm{BF}}$ (eV) & $\xi_\mathrm{F}$ (eV) \\
\hline
1999 Shimizu ~\cite{Shimizu1999JAP}   & Au   & 0.01 & 1.70 & -0.08 \\
2024 Buzio ~\cite{Buzio2024JoPDAP} & Au   & 0.01 & 1.31 & -0.08 \\
\hline
2024 Buzio ~\cite{Buzio2024JoPDAP} & Pt   & 0.01 & 0.92 & -0.04 \\
2009 Kim ~\cite{Kim2009JoPDAP}     & Pt   & 0.1  & 0.82 & -0.65 \\
\hline
2004 Postma ~\cite{Postma2004JAP}      & LSMO & 0.01 & 0.97 & 0.38 \\
2013 Rana ~\cite{Rana2013PRB}          & LSMO & 0.01 & 0.87 & 0.07 \\
2004 Postma ~\cite{Postma2004JAP}      & LSMO & 0.1  & 0.78 & 0.18 \\
2007 Ruotolo ~\cite{Ruotolo2007PRB}    & LSMO & 1    & 1.55 & -0.07 \\
\hline
1998 Yamamoto ~\cite{Yamamoto1998JJAP} & BKBO & 0.01 & 1.97 & -0.05 \\
1998 Yamamoto ~\cite{Yamamoto1998JJAP} & BKBO & 0.05 & 2.34 & -0.01 \\
1998 Yamamoto ~\cite{Yamamoto1998JJAP} & BKBO & 0.5  & 2.43 & -0.03 \\
\end{tabular}
\end{ruledtabular}
\end{table}

\subsubsection{Quantification of Interfacial Layer}
\label{subsec:IIA2}

To quantify the interfacial layer, an early work by Shimizu \textit{et al}. defined $\delta$/$\epsilon_\mathrm{i}$ to represent an effective vacuum-equivalent thickness,~\cite{Shimizu1999JAP} with  $\delta$ and $\epsilon_\mathrm{i}$ being thickness and the permittivity of the interfacial layer, respectively.~\cite{Ng2007} A high quality interface corresponds to a smaller $\delta$/$\epsilon_\mathrm{i}$, accompanied by smaller I-V hysteresis.~\cite{Mikheev2014NatureCommunications}  Quantitative analysis of Au/Nb:STO junctions yielded an averaged $\delta$/$\epsilon_\mathrm{i}$  value of 10.4.~\cite{Shimizu1999JAP} When $\epsilon_i=\epsilon_0$, the $\delta$/$\epsilon_\mathrm{i}$ value of 10.4 $V$m$^2$/C corresponds to 9.2$\cdot 10^{-2}$~nm thickness of the intrinsic interface layer.~\cite{Shimizu1999JAP} This parameter is crucial because the magnitude of RS directly scales with $\delta$ and RS tends to be suppressed when $\delta$/$\epsilon_\mathrm{i}$ is sufficiently small.~\cite{Mikheev2014NatureCommunications}

The ideality factor, $n$, is another critical parameter used to quantify the presence and influence of the interfacial layer and is extracted from the forward bias region of the I-V characteristics based on the thermionic emission model:
\begin{equation}
       I = S A^{*} T^{2}
\exp\!\left(-\frac{q\phi_{B}}{k_{\mathrm{B}}T}\right)
\left[
\exp\!\left(\frac{qV}{n k_{\mathrm{B}}T}\right)-1
\right],
    \label{eqn:3}
\end{equation}
 where $S$ is the junction area, $A^*$ is the Richardson constant, $\phi_B$ is the SBH, $n$ is the ideality factor, $V$ is the applied voltage, $q$ is the electron charge, $k_B$ is the Boltzmann constant, and $T$ is temperature.~\cite{Ng2007}
A value of $n = 1$ conventionally indicates pure thermionic conduction. For M/Nb:STO junctions, ideality factors are frequently reported significantly higher than 1, often ranging between $n=1.5$ to $n=3$.~\cite{Park2008JAP} There are two common origins for $n>1$, including electron tunneling through the barrier~\cite{Susaki2007PRB,Padovani1966SSE,Ramadan2005PRB} and voltage-dependent barrier height, which is often linked to the presence of an insulating interfacial layer and/or surface states.~\cite{Mikheev2014NatureCommunications, Bourim2014JSSST,Turut1996PS}

\begin{figure}
    \centering
    \includegraphics[width=\linewidth]{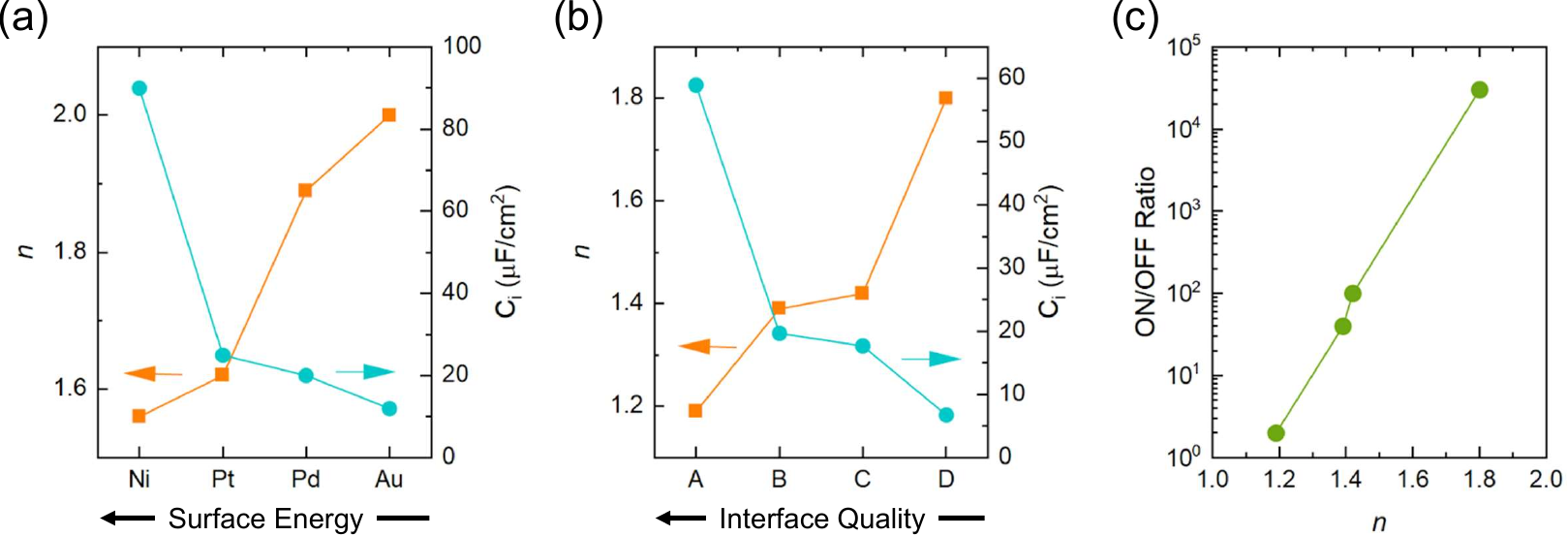}
    \caption{(a) Correlation between $n$ and $C_\mathrm{i}$ for different metals deposited with e-beam evaporation, and the arrow indicates increasing surface energy. (b) Correlation between $n$ and $C_\mathrm{i}$ for different deposition conditions of Pt: sample A was sputtered at 825~$^\circ$C after a two hour anneal at 825~$^\circ$C; sample B was sputtered at room-temperature after a two hour anneal at 825~$^\circ$C; sample C was sputtered at room temperature after a five minute anneal at 120~$^\circ$C; and sample D was E-beam evaporated with no annealing.~\cite{Mikheev2014NatureCommunications} Interface quality increases from sample D to A. (c) Correlation between the RS performance and $n$ for different growth condition of Pt/Nb:STO. Data for (a) was digitized from Park \textit{et al}.~\cite{Park2008JAP} and (b, c) from Mikheev \textit{et al}.~\cite{Mikheev2014NatureCommunications}.}
    \label{fig:8}
\end{figure}

Within the MIS electrostatic framework, the relationship between $n$ and the interface capacitance layer, $C\mathrm{_i}$, is defined as:\cite{Mikheev2014NatureCommunications}
\begin{equation}
    n = 1 + \frac{C\mathrm{_d}}{C\mathrm{_i}} = 1 + \frac{\mathrm{\delta}}{\mathrm{\epsilon_i}}\cdot \frac{\mathrm{\epsilon_s}}{\mathrm{W_d}},
    \label{eqn:4}
\end{equation}
where $\mathrm{\delta}$ and $\mathrm{W_d}$ are the interface and depletion layer thicknesses, and $\mathrm{\epsilon_i}$ and $\mathrm{\epsilon_s}$ are their dielectric constants, and $C_i$ and $C_d$ are their capacitance, respectively. It can be seen that $n$ is proportional with $\delta$/$\epsilon_\mathrm{i}$, which is the effective vacuum-equivalent thickness defined by Shimizu \textit{et al}., and inversely proportional to $C_\mathrm{i}$ (i.e., $C_\mathrm{i} \propto \frac{\epsilon_\mathrm{i}}{\delta}$).~\cite{Shimizu1999JAP, Park2008JAP} Therefore, $n$ can be considered a straightforward parameter to evaluate the interface quality in the Schottky junction. Fig.~\ref{fig:8}a and b show the junction parameters $n$ and $C_\mathrm{i}$, where junctions with larger $C_\mathrm{i}$ values possess smaller $n$ for different metals and deposition conditions.~\cite{Park2008JAP, Mikheev2014NatureCommunications} It is interesting to find out the Au/Nb:STO interface quality is the lowest among the plotted metals, agreeing with the Pt/Nb:STO interface being more robust than Au/Nb:STO.~\cite{Buzio2024JoPDAP} The trend of $n$ correlates well with the calculated surface energies of Ni, Pt, Pd, and Au, where Ni has the largest surface energy and Au the lowest.~\cite{Jian2004CP} The higher surface energy may favor improved wetting and interfacial bonding, consistent with the observed trend toward better interface quality and more homogeneous SBH distribution. For epitaxial Pt/Nb:STO, $n \approx 1.2$ and the ON/OFF ratio drops to roughly 2. (Fig.~\ref{fig:8}c). In fact, the interface quality directly determines the I-V hysteresis and ON/OFF ratio. Fig~\ref{fig:8}c shows the direct correlation between $n$ and ON/OFF ratios of a Pt/Nb:STO junction. This is consistent with the suppression of RS in PLD deposited (Pt, Au)/Nb:STO, where Pt/Au has intimate contact with Nb:STO, thus suppressing $\delta$ and RS.~\cite{Buzio2024JoPDAP}

\begin{figure}
    \centering
    \includegraphics[width=\linewidth]{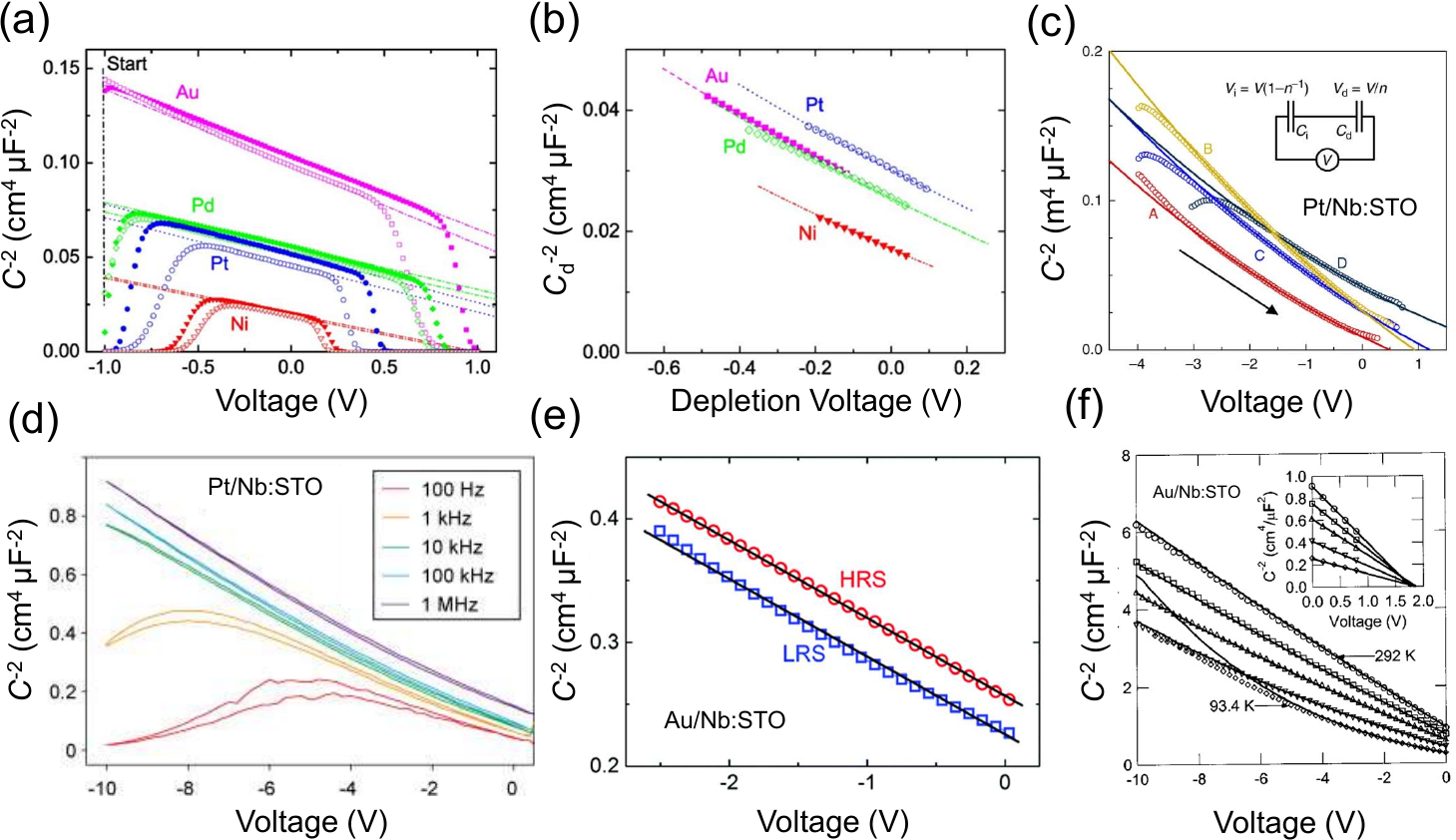}
    \caption{(a) $C^{-2}$ vs. $V$ for various M/Nb:STO junctions and (b) $C_\mathrm{d}^{-2}$ plotted versus the depletion voltage. (c) Non-linear $C^{-2}$ vs. $V$ measurements on Pt/Nb:STO for various metallization conditions, where A$\rightarrow$D indicates worsening interface quality. (d) Frequency-dependence of Pt/Nb:STO $C^{-2}$ vs. $V$  curves. (e) Resistance state dependence of Au/Nb:STO $C^{-2}$ vs. $V$  curves taken at 400~K. (f) Temperature-dependence of Au/Nb:STO $C^{-2}$ vs. $V$ curves from 292 to 93~K, and the inset displays the $x$-intercepts. Figues (a, b) taken from Ref.~\cite{Park2008JAP}, (c) from Ref.~\cite{Mikheev2014NatureCommunications}, (d) from Ref.~\cite{Li2010MSEB}, (e) from Ref.~\cite{Fan2017JoMCC}, and (f) from Ref.~\cite{Shimizu1999JAP}.}
    \label{fig:9}
\end{figure}

Capacitance vs. voltage (C-V) characteristics have been often used to understand the SBH and $W_\mathrm{d}$, since the built-in potential $V_\mathrm{bi} = \phi_\mathrm{B} + \xi_F$ and $C \propto \epsilon_s/W_\mathrm{d}$. In an ideal Schottky diode, the relationship between the depletion capacitance and voltage in the reverse bias region is given by:
\begin{equation}
    C^{-2} = \frac{2}{q\epsilon_S N_D A^2}(V_{bi} - V - \frac{k_BT}{q}),
    \label{eqn:5}
\end{equation}
where $V_{bi}$ is the built-in potential, $q$ is the elementary charge, $\epsilon_s$ is the semiconductor permittivity, $N_D$ is the donor concentration, and the $A$ is diode area.~\cite{Park2008JAP, Susaki2007PRB} Eq. 5 assumes there are no contributions from interfacial layer, series capacitors, and interface traps, and overestimates $V_\mathrm{bi}$ for various metal: Au = 2.65 eV, Pd = 2.55 eV, and Pt = 2.06 eV (see Fig.~\ref{fig:9}a).~\cite{Park2008JAP} As discussed in Section~\ref{subsec:IIA1}, M/Nb:STO junctions deviate from the ideal Schottky model and behave as a MIS structure due to the existence of an interfacial layer. For example, Yoshida \textit{et al}. explained the lower $C$ and higher $V_{bi}$ of Au/Nb:STO by assuming a low dielectric constant interfacial layer.~\cite{Yoshida1991JournalofAppliedPhysics} When an interfacial layer is present, the measured capacitance is a series combination of $C_i$ and $C_d$. The relationship between junction capacitance and the applied voltage is re-defined as:~\cite{Yang2014JAP}

\begin{equation}
    C^{-2} = \frac{2n}{q\epsilon_s N_D A^2}(nV_{bi} - V - \frac{nk_BT}{q}).
    \label{eqn:6}
\end{equation}
It can be seen that Eqn.~\ref{eqn:6} is very similar to the standard Eqn.~\ref{eqn:5} except for the appearance of $n$ to account for the interfacial layer, giving the effective built-in potential of $n \cdot V_\mathrm{bi}$.~\cite{Mikheev2014NatureCommunications} Although $C^{-2}$ is still a linear function of $V$, $V_\mathrm{bi}$ is overestimated by $n$ if one still uses Eqn.~\ref{eqn:5}. Alternatively, Park \textit{et al}. extracted from the true $V_\mathrm{bi}$ (i.e., the voltage drop across the depletion layer, $V_\mathrm{d}$) through a careful selection of $C_\mathrm{i}$ and fitting $C_\mathrm{d}^{-2}$ vs. $V_\mathrm{d}$ as seen in Fig.~\ref{fig:9}b. These methods provide the necessary corrections for M/Nb:STO to acquire accurate $V_\mathrm{bi}$ values from capacitance measurements. 

Two common traits of M/Nb:STO $C^{-2}$ vs. $V$ curves are the presence of hysteresis (Fig.~\ref{fig:9}a and d) and non-linear voltage dependence (Fig.~\ref{fig:9}c and d). The C-V hysteresis originates from the barrier profile and interface chemistry shifting with applied voltage, such as charge trapping/detrapping.~\cite{Park2008JAP} This is even observed at frequencies as large as 1 MHz.~\cite{Park2008JAP} The hysteresis becomes more pronounced at low frequency (Fig.~\ref{fig:9}d) and low $T$.~\cite{Li2010MSEB} This is consistent with slower equilibration rates at low $T$.
On a related note to C-V hysteresis, Fan \textit{et al}. found that the HRS and LRS of Au/Nb:STO possessed different $V_\mathrm{bi}$ values of 1.78 and 1.55 eV, respectively (Fig.~\ref{fig:9}e).~\cite{Fan2017JoMCC} This provides support for uniform barrier modulation. Such a small C-V hysteresis only exists when detrapping is incomplete (e.g., voltage measurement range is small). In fact, LRS is largely similar to HRS in this case. For a I-V hysteresis like Fig.~\ref{fig:5}c, only HRS gives reasonable C-V curve while LRS can not give true C-V due to high leakage current. Another salient feature is the temperature- and field-dependent permittivity of STO, $\epsilon_s(E,~T)$,~\cite{Shimizu1999JAP, Susaki2007PRB} which commonly gives non-linear $C^{-2}$ vs. $V$. This nonlinearity becomes more pronounced at low $T$, even with a nearly $T$-independent $V_\mathrm{bi}$ (Fig.~\ref{fig:9}f).~\cite{Shimizu1999JAP} Nonlinear features may also arise from current leakage, and C-V measurements are typically taken assuming a capacitor-resistor parallel circuit.~\cite{Fujii2007PRB,Li2010MSEB, Yang2014JAP} More information on non-linear fittings can be found in Shimizu \textit{et al}.,~\cite{Shimizu1999JAP} Suzuki \textit{et al}.~\cite{Suzuki1997JAP}, Mikheev \textit{et al},~\cite{Mikheev2014NatureCommunications} and Kim \textit{et al}.~\cite{Kim2020APL}

On a related note, it is likely that spatial inhomogeneities in the SBH also occur,~\cite{Kan2013APL, Lee2011APL,Park2008JAP} given by "patches" of OV-rich concentration, terrace step-edges, process-induced damage, and contamination, which promote active tunneling areas while the areal majority is not degraded.~\cite{Tung1992PRB} This is related to a more conventional approach to $n$, where increased barrier inhomogeneities lead to a larger $n$.~\cite{Werner1991JAP,Tung1992PRB,Tung2001MSER} This can be captured by the $T$-dependence of $n$ according to this linear relationship:
\begin{equation}
    n^{-1} - 1= A  + \frac{B}{2k_\mathrm{B}T},
\end{equation}
where $A$ captures the sensitivity of $\phi_\mathrm{B}$ to electric fields and $B$ captures the standard deviation of $\phi_\mathrm{B}$.~\cite{Hikita2008PRB,Werner1991JAP} While this requires $T$-dependence, it provides a useful measure of spatial homogeneity without invoking conducting atomic force microscopy (CAFM) or scanning tunneling microscopy (STM) measurements, in addition to capturing the voltage sensitivity which underpins the MIS model. 

\subsubsection{Physical Origins of the Intrinsic and Extrinsic Interfacial Layer}
As discussed in previous sections, an effective electrostatic interfacial layer exists between M and Nb:STO. The physical origin of this interfacial layer could be related to surface or interface imperfections, such as electrode bonding, step terraces, dangling/damaged bonds, and other structural imperfections, as O$_2$ annealing and O$_3$ treatment cannot completely eliminate such a layer.~\cite{Shimizu1995APL, Shimizu1999JAP, Yamamoto1998JJAP, Buzio2012APL, Mikheev2014NatureCommunications,Rodenbucher2013NJP} Often such relatively clean samples with $n$ close to unity exhibit no or very limited RS. ~\cite{Buzio2024JoPDAP,Mikheev2014NatureCommunications} Fig.~\ref{fig:8}c clearly shows that higher $n$ leads to larger I-V hysteresis and ON/OFF ratios. 
While electrodes in intimate metal–oxide contact, such as epitaxial heterostructures, are expected to exhibit minimal interfacial layers,~\cite{Ng2007} these heterostructures still display a low-temperature anomaly in the dielectric permittivity.~\cite{Yoshida1991JournalofAppliedPhysics, Yoshida1991PhysicaB, Susaki2007PRB} This behavior has been attributed to an intrinsic interfacial layer arising from the electronic and structural properties of the metal–oxide interface. In contrast, samples with an extrinsic interfacial layer, originating from poor metal–oxide contact,~\cite{Ng2007} exhibit $n \gg 1$ and pronounced I–V hysteresis, consistent with significant interfacial charge trapping.~\cite{Mikheev2014NatureCommunications}

Some earlier works reported that RS can be impacted by surface treatments~\cite{Li2010MSEB} and moisture.~\cite{Kawada2004JE} The aging effect may give a clue to the origin of such an interfacial layer.~\cite{Ohsawa2021JoPCC,Hirose2019JAP} Hirose \textit{et al}. found that freshly fabricated Au/Nb:STO junctions are characterized by extremely small rectification and negligible I-V hysteresis loop, while a week of exposure in air induced a large I-V hysteresis loop.~\cite{Hirose2019JAP} Interestingly, samples stored in vacuum and Ar did not exhibit such aging behavior.~\cite{Hirose2019JAP}. This result indicates that the interaction between air and the sample promotes the extrinsic interfacial layer formation. Many papers suspect that environmental oxygen could be the dominating factor for such aging effect and consequently the RS.~\cite{Bian2025FML,Ke2011APL,Andreasson2009APL,Goux2010APL,Buzio2012APL,Ohsawa2021JoPCC,Hirose2019JAP}

\begin{figure}
    \centering
    \includegraphics[width=0.9\linewidth]{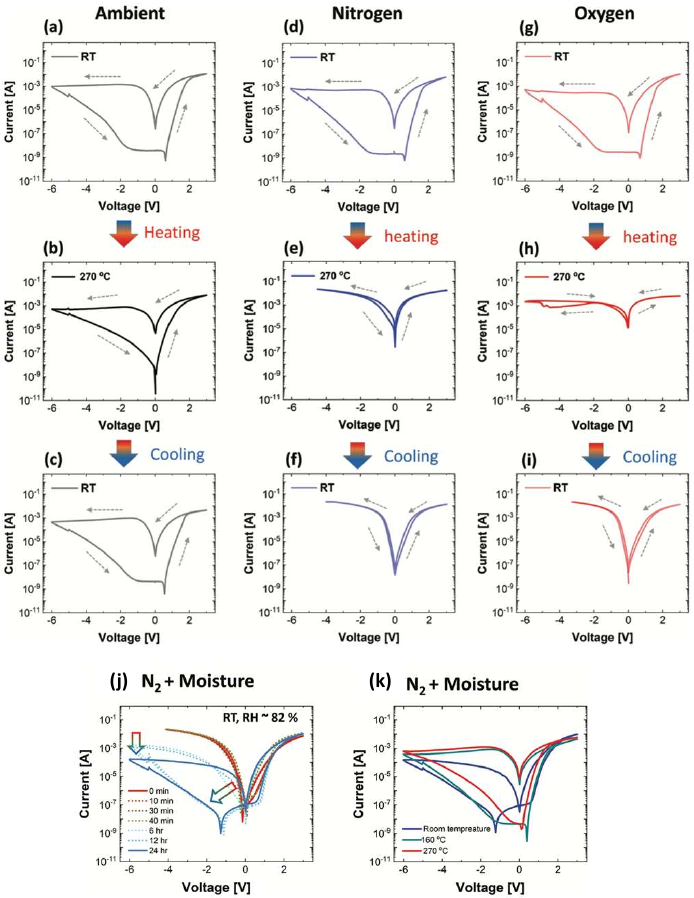}
    \caption{(a-c) RS dependence of Au/Nb:STO in air (a) at room-temperature, (b) after heating to 270 $^\circ$C, and (c) after cooling back to room temperature. The same heating procedure for (d-f) nitrogen and (g-i) oxygen. (j) Aging dependence and (k) temperature dependence of RS in humid nitrogen. Figures taken from Ref.~\cite{Kunwar2023AEM}. }
    \label{fig:10}
\end{figure}

Our recent work claimed that water moisture in air plays a critical role in RS and is responsible for the extrinsic layer formation. To understand what components in air facilitate the interfacial layer formation, Kunwar \textit{et al}. measured Au/Nb:STO junctions in controlled environments and various temperatures.~\cite{Kunwar2023AEM} Three similar samples have been heated from room temperature to 270~$^\circ$C and back down to room temperature in air, dry N$_2$, and dry O$_2$, separately.~\cite{Kunwar2023AEM} I-V curves were collected at these temperatures and conditions. The samples show similar I-V hysteresis loops with a large ON/OFF ratio at room-$T$ before heating.~(Fig.~\ref{fig:10}a, d, and g) When heated to 270~$^\circ$C, the samples in air showed suppressed I-V, mostly due to the reduced HRS~(Fig.~\ref{fig:10}b). The I-V hysteresis loops for samples heated to the same temperature in dry N$_2$ and dry O$_2$, however, are completely collapsed.~(Fig.~\ref{fig:10}e and h) More specifically, HRS collapsed towards the LRS. After cooled to room temperature, the sample in air restores its original I-V hysteresis loop.~(Fig.~\ref{fig:10}c) Strikingly, I-V loops for these two samples processed in dry N$_2$ and dry O$_2$ keep collapsed states~(Fig.~\ref{fig:10}f and i). The I-V curves for these two samples are nearly unchanged even after 24 hours.~\cite{Kunwar2023AEM} Once exposed to wet N$_2$, I-V loops evolve with time and hysteresis starts to develop with obvious ON/OFF ratio after exposure to moisture for 24 hours at room temperature.~(Fig.~\ref{fig:10}j) More strikingly, the original I-V loop with huge hysteresis can be restored after heating this sample to 160~$^\circ$C and 270~$^\circ$C~(Fig.~\ref{fig:10}k). This work confirms that moisture could dominate the interfacial layer formation, since N$_2$ and O$_2$ environments suppress RS, and rationalizes previous observations that the formation of the extrinsic interfacial layer is correlated with electrode quality and aging effect in air.~\cite{Hirose2019JAP, Park2008JAP, Mikheev2014NatureCommunications, Buzio2024JoPDAP} In fact, the impact of water moisture on RS has been discussed in other systems.~\cite{Wang2023FP} For example, F. Messerschmitt \textit{et al}. systematically studied the effect of moisture in Pt/SrTiO$_\mathrm{3-x}$/Pt junctions.~\cite{Messerschmitt2015AFM,Messerschmitt201AEM}


The extrinsic interfacial layer likely hosts electron traps to modulate the Schottky barrier. ~\cite{Mikheev2014NatureCommunications, Fan2017JoMCC} Electron traps are localized energy states near the Fermi energy that carriers can scatter into, becoming localized until re-emitted into a conduction band. Charged defects create these states by introducing donor-like levels, allowing them to capture electrons and act as dynamic trapping centers, whose population can evolve with voltage bias, temperature, or ionic motion.~\cite{Park2014APL, Lin2013PRL,Waser2007NM, Yang2013NN} Given the experimental work by Kunwar \textit{et al.},~\cite{Kunwar2023AEM}, there is strong evidence that protonic defects underpin charge trapping in the extrinsic interfacial layer. The protonation scenario, which is essentially described by the water-splitting effect in STO, is a multi-step process for the incorporation of moisture and production of hydrogen gas.~\cite{Ogawa2025JACS,Domen1980JACSCC,Wagner1980JACS}  A first order effect is a water molecule (H$_2$O) interacting with a neutral oxygen ($\mathrm{O}^\times _\mathrm{O}$) and a doubly charged OV ($V_\mathrm{O}^{\bullet\bullet}$), as described by the following equation:
\begin{equation}
    \mathrm{H_2O}(g) + \mathrm{O}_\mathrm{O}^{\times} + V_\mathrm{O}^{\bullet\bullet} \rightleftharpoons \mathrm{2OH^\bullet _O},
\end{equation}
where the resulting trapping state is a hydroxide molecule inhabiting an oxygen site ($\mathrm{OH^\bullet _O}$),~\cite{Kessel2010APL, Messerschmitt2015AFM} i.e., a protonic defect.~\cite{Lubben2018AEM} The incorporation of moisture into the lattice may either lead to $\mathrm{O}_\mathrm{O}^{\times}$ or $\mathrm{OH^-}$ migrating towards the interface, as described by the following equations:~\cite{Kessel2010APL, Messerschmitt2015AFM}
\begin{equation}
    \mathrm{2OH^\bullet _O} + \mathrm{2e^-} \rightleftharpoons 2\mathrm{O}_\mathrm{O}^{\times} + \mathrm{H_2}(g)
\end{equation}

\begin{equation}
    \mathrm{2H_2O}(g) + \mathrm{2e^-} \rightleftharpoons \mathrm{2OH^-} + \mathrm{H_2}(g).
\end{equation}
These are second order effects of protonic defects in the lattice, where the surface chemistry includes a mixture of oxygen atoms and hydroxyl molecules. OVs have also been widely discussed as a possible trapping site in M/Nb:STO,~\cite{Chen2011APL, Bourim2014JSSST, Shen2013APA, Park2014APL, Buzio2012APL,Quinonez2025APLED, Zhong2013CAP,Song2026MSSP} due to their ubiquitous nature and role in conventional oxide switching layers.~\cite{Cui2013AMI, Hong2009CPL, Zhu2012JoPDAP,Srivastava2017AMI,Zeumault2022JAP,Peng2023AOM, Liao2024APL,Gao2026AFM} While native lattice defects, such as OVs, have traditionally been considered the primary trapping sites,~\cite{Ni2007APL,Hao2015PRB} recent evidence highlighting the critical role of moisture and protonic defects underscores the need for new studies that directly compare the respective contributions of native lattice and protonic defects to trapping behavior.


\subsection{Schottky Barrier Modulation by Electron Trapping}
\subsubsection{MIS Model Coupled with Traps}
\label{sec:IIA4}
\begin{figure}
    \centering
    \includegraphics[width=0.4\linewidth]{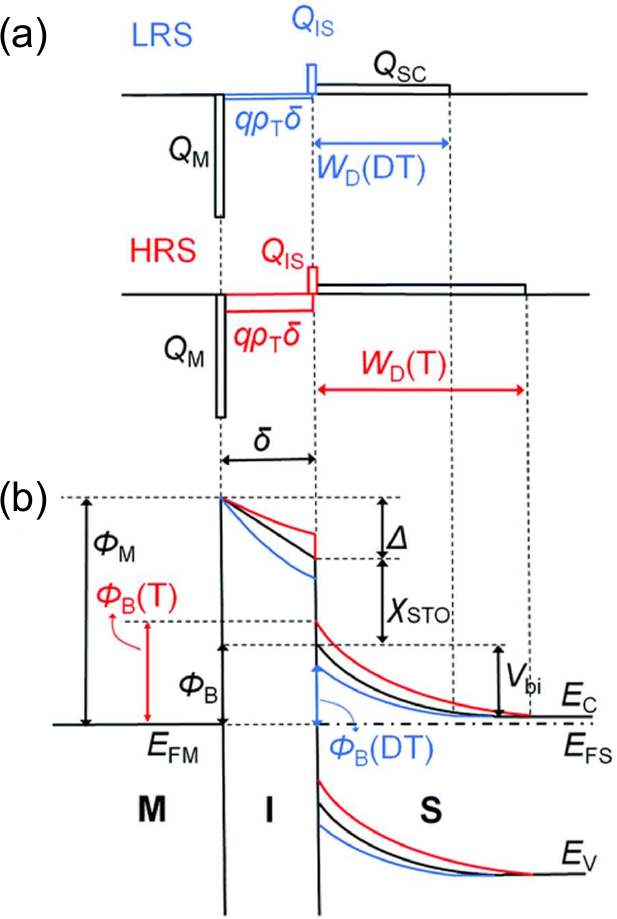}
    \caption{(a) The impact of trapped charges, interface states, and space-charge on the LRS via detrapping (DT) and HRS via trapping (T). (b) The position dependent energy diagram for the MIS model for the DT and T modulation. Figure taken from Ref.~\cite{Fan2017JoMCC}.}
    \label{fig:11}
\end{figure}

The electronic MIS model discussed in Sec.~\ref{subsec:IIA1} and \ref{subsec:IIA2} well describes the I-V and C-V results in M/Nb:STO junctions, and making the extension to traps coupled with the MIS model can explain the RS behavior and its dynamics. Several publications discussed such physical processes, including in Pt/Nb:STO,~\cite{Mikheev2014NatureCommunications, Bourim2014JSSST} Au/Nb:STO,~\cite{Fan2017JoMCC, Shen2013APA} and Ni/Nb:STO.~\cite{Yin2015PCCP, Goossens2018JAP} The charge trapping and detrapping at the interface modulates the Schottky barrier, conduction, and consequently the ON/OFF ratio. The physical process of how charge trapping/detrapping modulates the SBH and $W_D$ is shown in Fig.~\ref{fig:11}. In short,  an applied forward bias  extracts electrons from the interface states (i.e., detrapping process), resulting in a loss of trapped charge and reduction of the SBH and $W_D$.~\cite{Mikheev2014NatureCommunications, Fan2017JoMCC} This evolves the junction into the LRS. Conversely, a reverse bias promotes the electron trapping at the interface, which increases the SBH and $W_D$, thus leading to the HRS.~\cite{Mikheev2014NatureCommunications, Fan2017JoMCC} While the modulation of the SBH is quite intuitive, the modulation of $W_\mathrm{d}$ originates from charge compensation between space-charge and interface charge.~\cite{Mikheev2014NatureCommunications} Thus, the charge trapping and detrapping mechanism impacts both thermionic emission and tunneling. 

As shown in Fig.~\ref{fig:11}, the \textit{ideal} SBH (i.e., no interfacial dipoles or pinning) is influenced by the interface charge, given by the equation: 
\begin{equation}
    \phi_\mathrm{B} = \phi_\mathrm{M} -\chi_\mathrm{S} - \Delta_\mathrm{C},
    \label{eqn:7}
\end{equation}
where $\phi_\mathrm{M}$ is the metal work function, $\chi_\mathrm{S}$ is the electron affinity of STO, and $\Delta_\mathrm{C}$ is the summation of energy associated with the space-charge (Q$_\mathrm{SC}$), trapped charge ($q \delta \rho_\mathrm{T}$), and interface states (Q$_\mathrm{IS}$).~\cite{Mikheev2014NatureCommunications, Fan2017JoMCC} These contributions to $\Delta_\mathrm{C}$ is given by:
\begin{equation}
    \Delta_\mathrm{C} = \frac{\delta}{\epsilon_\mathrm{i}\epsilon_0}Q_\mathrm{SC} + \frac{\delta^2}{2\epsilon_\mathrm{i}\epsilon_0}q\rho_\mathrm{T} + \frac{\delta}{\epsilon_\mathrm{i}\epsilon_0}Q_\mathrm{IS},
    \label{eqn:8}
\end{equation}
where $\rho_\mathrm{T}$ is the trapped charge density and other related terms are shown in Fig.~\ref{fig:11}.~\cite{Fan2017JoMCC}  $\Delta_\mathrm{c}$ can be calculated using an electrostatic model by solving Poisson's equation for a piecewise spatial charge distribution. This definition of $\Delta_\mathrm{C}$ follows Fan \textit{et al}.'s derivation;~\cite{Fan2017JoMCC} however, Mikheev \textit{et al}. used a centroid charge instead of a homogeneous spatial distribution.~\cite{Mikheev2014NatureCommunications} Each method provides similar results. Eqns.~\ref{eqn:7} and \ref{eqn:8} lead to the following equation for the SBH:
\begin{equation}
    \phi_\mathrm{B} = (\phi_\mathrm{M}-\chi_\mathrm{STO}) - \frac{\delta}{\epsilon_\mathrm{i}\epsilon_0}Q_\mathrm{SC} - \frac{\delta^2}{2\epsilon_\mathrm{i}\epsilon_0}q\rho_\mathrm{T} - \frac{\delta}{\epsilon_\mathrm{i}\epsilon_0}Q_\mathrm{IS}.
    \label{eqn:9}
\end{equation}
These values can be extracted from conventional electronic characterizations: $\phi_\mathrm{B}$ can be extracted from thermionic emission I-V fittings; $\phi_\mathrm{M}-\chi_\mathrm{STO}$ remains constant (i.e., $\phi_\mathrm{M} = 5.65$~eV for Pt and $\chi_\mathrm{STO} = 3.9$~eV); $\frac{\delta}{\epsilon_\mathrm{i}\epsilon_0}Q_\mathrm{SC}$ can be extracted from C-V, since $Q_\mathrm{SC} = qN_\mathrm{D}W_\mathrm{d}$ and $C_\mathrm{i} = \frac{\epsilon_\mathrm{i}\epsilon_0}{\delta}$; and the trapped charge contribution, Q$_\mathrm{T}=q\delta \rho_\mathrm{T}+$Q$_\mathrm{IS}$, can then be calculated from Eqn.~\ref{eqn:9}.~\cite{Mikheev2014NatureCommunications} $Q_\mathrm{T}$ can be positive or negative, depending upon the relative magnitude of positive interface states, $Q_\mathrm{IS}$, and negative trapped charge, $q \delta \rho_\mathrm{T}$, within the interface layer.~\cite{Mikheev2014NatureCommunications,Fan2017JoMCC} This phenomenon is a function of interface thickness, $\delta$, leading to negative $Q_\mathrm{T}$ as $\delta$ increases.~\cite{Mikheev2014NatureCommunications} 


If intrinsic $Q_\mathrm{IS}$ does not change significantly during switching and the $Q_\mathrm{SC}$ term changes weakly compared to the $\rho_\mathrm{T}$ term (i.e., large $\delta$), the SBH change is determined by the electrostatic potential drop across the interfacial layer.
While the semiconductor space charge adjusts to compensate the trapped charge,~\cite{Mikheev2014NatureCommunications} the dominant contribution arises from charge trapping/detrapping through $\rho_\mathrm{T}$, leading to a SBH change:
\begin{equation}
    \Delta \phi_\mathrm{B} = -\frac{\delta^2}{2\epsilon_\mathrm{i}\epsilon_0} q \Delta \rho_\mathrm{T}.
\end{equation}
Since the interface trapped charge possesses a negative sign,~\cite{Fan2017JoMCC} an increase in $\rho_T$ results in an increased SBH. This represents a change in volume charge density in the interfacial layer (i.e., C/m$^3$). If $q\rho_\mathrm{T}$ is approximately uniform across $\delta$, then the change in trapped charge is:$\Delta Q_\mathrm{T} \approx q \delta \Delta \rho_\mathrm{T}$. Thus, the SBH change becomes: 
\begin{equation}
    \Delta \phi_\mathrm{B} \approx - \frac{\Delta Q_\mathrm{T}}{2C_\mathrm{i}}.
\end{equation}
This is the SBH change via trapped charge, and switching is electrostatic and reversible, governed by trap occupancy. This can be understood in the following limits: if $C_\mathrm{i}$ is large (i.e., $\delta$ is small) then $\Delta \phi_\mathrm{B}$ is small, and if $C_\mathrm{i}$ is small (i.e., $\delta$ is large) then $\Delta \phi_\mathrm{B}$ is large. This further supports that $\delta$ is the primary indicator of RS performance in M/Nb:STO. 



The negative trapped charge, which is an effect of reduced interface quality (i.e., larger $\delta$), explains both the $Q_\mathrm{T}$ crossover and the linear scaling of $W_\mathrm{d}$ with $\delta$.~\cite{Mikheev2014NatureCommunications} 
From various reports, it can be corroborated that the intrinsic interface between M/Nb:STO does not yield resistive switching as the device is always in the LRS when there is no trapped charge.~\cite{Mikheev2014NatureCommunications, Buzio2024JoPDAP, Park2014APL,Goossens2018JAP} Further, this mechanism agrees with the fact that some devices tend to show RS after allowing them to develop an interface charge layer (i.e., HRS) if they do not show RS in pristine or as-fabricated conditions.~\cite{Hirose2019JAP, Kunwar2023AEM, Ohsawa2021JoPCC} Overall, this demonstrates that RS in M/Nb:STO is not intrinsic, but instead emerges from the development of an extrinsic interfacial layer.


\subsubsection{Signature of Charge Trapping/Detrapping}

There are several possible origins of these trap states, which are generally associated with defects, such as oxygen vacancies (OVs), hydroxyl species (e.g., OH), fabrication process induced damage, and unintentional contamination.~\cite{ Shimizu1997ASS,Zhu2005AM, Son2024JKPS, Kunwar2023AEM} OVs act as native defects and electron donor dopants within the Nb:STO material, concentrating near the interface.~\cite{Hamid2009APA,Pacchioni2003CPC, Lee2014APLM} Generally, native defects, such as Frenkel defects, interstitial defects, and vacancies, create trapping layers or interface states that are available to capture or release electrons under an external electric field.~\cite{Hamid2009APA,Cheng2021JoPCL, Park2014APL,Chen2011APL, Ni2007APL,Ohnishi2005APL,Choi2006APL} OV-assisted electron trapping/detrapping has been widely adopted to explain Schottky barrier modulation.~\cite{Brillson2011JAP,Dharanya2022JNP,Bourim2013CAP,Buzio2012APL,Bian2025FML,Bourim2014JSSST,Chen2011APL,Park2014APL,Lee2014APLM,Quinonez2025APLED,Shen2013APA,Zhong2013CAP,Chen2010APL} Another common argument for trapping sites is correlated with the processing conditions and unintentional contamination.~\cite{Son2024JKPS, Li2023APL, Mikheev2014NatureCommunications}  More recently, the introduction of moisture has shown to be an important factor in RS.~\cite{Kunwar2023AEM} Regardless, the role of surface and interface defects is profound in interface-dominated RS mechanisms. 

\begin{figure}
    \centering
    \includegraphics[width=\linewidth]{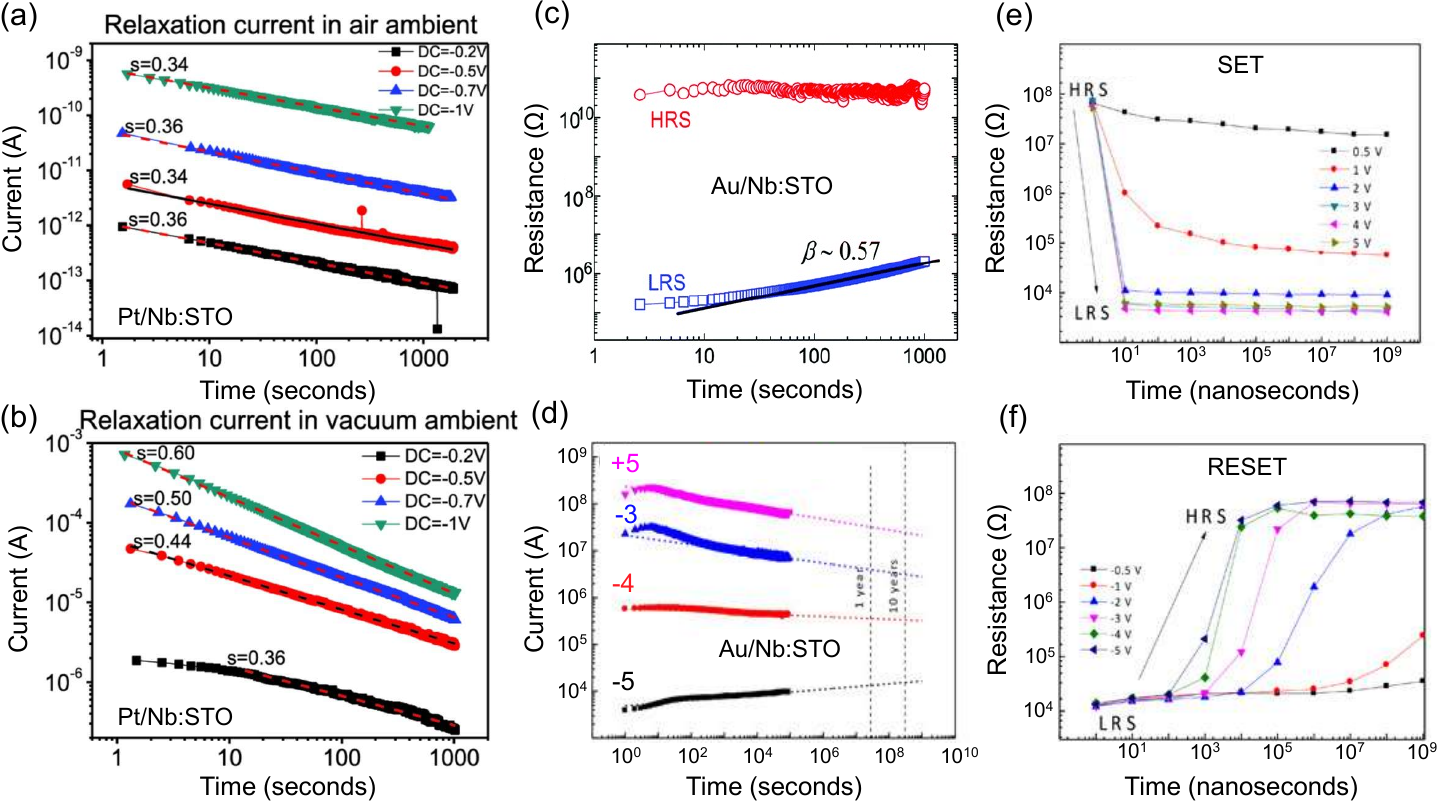}
    \caption{Read voltage dependent retention characteristics for e-beam evaporated Pt/Nb:STO in (a) air and (b) vacuum conditions using a set voltage of 2~$V$. (c) Retention curves of e-beam evaporated Au/Nb:STO for the HRS and LRS measured with a read voltage of -0.2~$V$ and SET (RESET) voltage of 3 $V$ (-6.5 $V$). (d) Long-term retention of various resistance states of sputtered Au/Nb:STO using a read voltage of -0.3~$V$ and a SET voltage of +5 and RESET voltages of -3, -4, and -5, where the power-law scaling exponent is 0.06, -0.03, -0.1, and -0.12 for the SET/RESET voltages of +5, -3, -4, and -5 $V$, respectively.~\cite{Li2018PSSA} Pulse-width and pulse-magnitude dependent switching for (e) SET and (f) RESET operation on sputtered Au/Nb:STO. Figures (a, b) taken from Ref.~\cite{Bourim2014JSSST}, (c) taken from Ref.~\cite{Fan2017JoMCC}, and  (d-f) from Ref.~\cite{Li2018PSSA}.}
    
    \label{fig:12}
\end{figure}

One of the key characteristics of the charge trapping/detrapping mechanism is that it shows strong resistance relaxation behavior. After applying a SET voltage pulse, the LRS often shows a power-law decay consistent with the Curie-von Schweidler law.~\cite{Tian2011APA,Bourim2014JSSST,Zhang2009APL, Goossens2018JAP, Mikheev2014NatureCommunications,Ni2007APL, Fan2017JoMCC, Shen2013APA, Mikheev2015SR,Kan2013APL} This behavior is a classic signature of charge recombination in disordered systems with a broad distribution of trap states. Such a LRS decay towards HRS has widely been reported when reading with a small negative voltage, in addition to a suppressed effect when reading with a small positive voltage (see Fig.~\ref{fig:12}a-d).~\cite{Tian2011APA,Bourim2014JSSST,Zhang2009APL, Goossens2018JAP, Mikheev2014NatureCommunications,Ni2007APL, Fan2017JoMCC,Kunwar2023AEM} When reading at relatively higher positive voltage after a reset pulse at negative bias often shows relaxation of HRS towards LRS.~\cite{Ni2007APL,Bourim2014JSSST,Goossens2018JAP, Shen2013APA, Yang2014JAP} More specifically, the positive reading voltage can detrap charges, which makes relaxation of HRS towards LRS. Despite the diffusion related retention characteristics, extrapolated retention of various resistance states give a relative change of approximately one order of magnitude over 10 years (Fig.~\ref{fig:12}d). The variable scaling exponents suggest that ambient conditions, metallization processes, and read voltage greatly influence the retention properties of M/Nb:STO. For example, Mikheev \textit{et al}. observed that increasing the M/Nb:STO interface quality decreased the decay exponent, i.e., better retention properties.~\cite{Mikheev2014NatureCommunications} 

On a related note, M/Nb:STO exhibits nanosecond RS dynamics, further supporting an electronic switching mechanism. Li \textit{et al}. found that the SET transition from HRS to LRS occurs at 10~ns with positive bias, while RESET transition from LRS to HRS occurs at $10^4 - 10^7$~ns under negative bias (see Fig.~\ref{fig:12}e and f).~\cite{Li2018PSSA} The authors proposed that the SET process is associated with electron detrapping, a relatively fast electronic effect, whereas electron trapping during the RESET process is limited by a slower ionic redistribution process involving OVs or other charged species.~\cite{Li2018PSSA}
In contrast, Zhang \textit{et al}. observed a more symmetric RS response and a strong dependence of the switching timescale on the electrode material, with a response time of $\sim$5~ns in Ag/Nb:STO compared with a response approximately two orders of magnitude slower in Pt/Nb:STO.~\cite{Zhang2010APL}
Further experimental and theoretical work is needed to fully understand the mechanisms underlying the ultrafast SET and RESET switching dynamics.



A common physical picture of RS is supported by a range of experimental techniques, including impedance spectroscopy, scanning probe microscopy, transport, and capacitance measurements. These experimental techniques have been implemented to study the trapped charge dynamics at the interfacial layer. For instance, admittance spectroscopy reveals that the characteristic lifetime of charge carrier at these traps is dependent on the resistance state (i.e., HRS or LRS) and the ambient environment.~\cite{Li2010MSEB,Kim2001JMR,Bourim2013CAP} This provides strong evidence that the surface potential and the electric field profile is being modified by charge dynamics at trap sites. Scanning Kelvin Prove Microscopy (SKPM) has been used to directly visualize changes in the surface potential after applying a bias with an AFM tip, supporting that charge can be trapped and released at the surface.~\cite{Wang2016ASS, Fan2017JoMCC} Electrical characterization frequently reveals hysteretic I-V and C-V curves, suggesting that both conductance and capacitance are modulated by the applied voltage history.~\cite{Park2008JAP} 
Together, this evidence links both the retention and switching to the dynamics of charge trapping/detrapping, which controls the evolution and stability of the resistance states. 

\subsection{Tunneling and Schottky Barrier Inhomogeneity}

\begin{figure}
    \centering
    \includegraphics[width=0.5\linewidth]{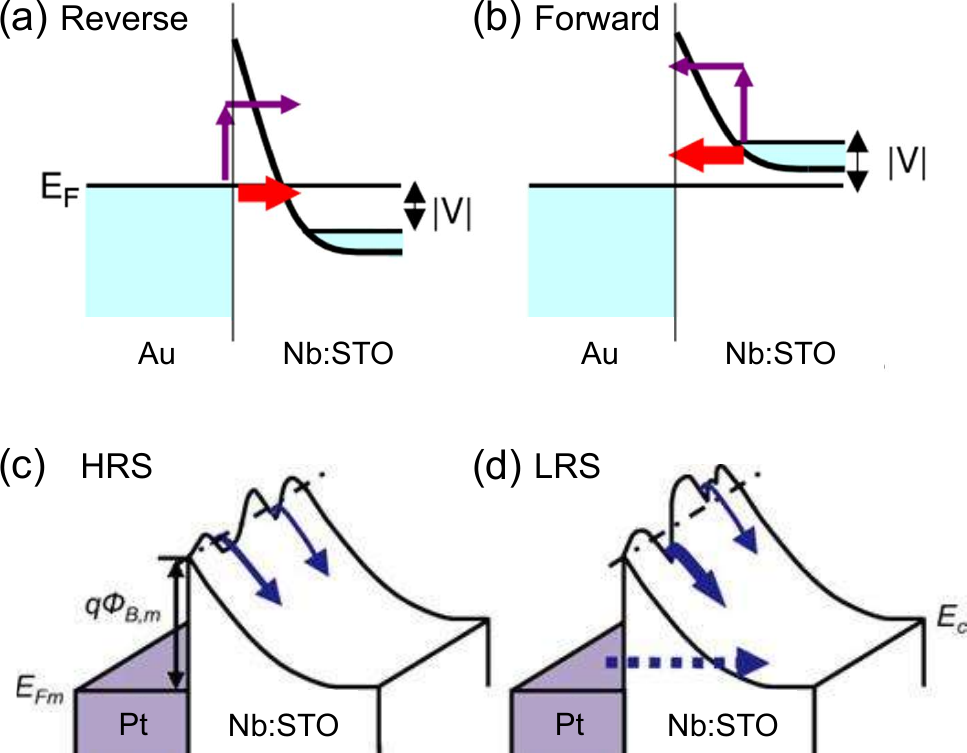}
    \caption{Diagram of tunneling through the Schottky barrier in the (a) reverse bias and (b) forward bias. The thin purple arrows describes thermionic field emission, while the thick red arrow field emission. Position dependent diagram of constant mean SBH ($q \phi_\mathrm{B, m}$) in the (c) HRS, with thermionic emission, and (d) LRS, with thermionic emission plus tunneling. Figures (a, b) taken from Ref.~\cite{Susaki2007PRB} and (c, d) taken from Ref.~\cite{Lee2011APL}. }
    \label{fig:13}
\end{figure}

Quantum tunneling represents a plausible transport channel that may contribute to, or even dominate, RS in M/Nb:STO, particularly in conjunction with Schottky barrier modulation. Low-$T$ and LRS transport can be dominated by tunneling due to the thinning of $W_\mathrm{d}$.  First, low-$T$ transport is considered, where suppression of thermal activation isolates field-driven processes and highlights $W_\mathrm{d}$ as the key tuning parameter. In this regime, transport can occur via field emission (FE), in which carriers can directly tunnel through the barrier from states near the Fermi energy assisted by an applied electric field (see red arrow in Fig.~\ref{fig:13}a and b), and thermionic field emission (TFE), where carriers are partially thermally excited above the Fermi level before for tunneling through a relatively thinner region of the barrier (see purple arrow in Fig.~\ref{fig:13}a and b).~\cite{Padovani1966SSE} The $T$-dependent scaling of these processes has demonstrated the dominance of TFE at room-$T$.~\cite{Hasegawa1991JAP, Susaki2007PRB} Next, trap-assisted tunneling (TAT)---a commonly proposed transport mechanism in oxide- and nitride-based Schottky junctions~\cite{Lim2015E,Kim2024TEEM,Al2020MRE, Mahaveer2006JAP}---is discussed, while noting its role remains uncertain and not yet established in M/Nb:STO. Finally, tunneling behavior is connected to RS characteristics, motivated by photoemission measurements showing that the SBH remains independent of the resistance state,~\cite{Shang2008APL,Lee2011APL} which could be consistent with barrier inhomogeneities dominating transport behavior~(Fig.~\ref{fig:13}c and d). These observations suggest the RS may not be governed by global changes in SBH, but by local modulation of tunneling pathways. 


\subsubsection{Tunneling at Low Temperatures}

\begin{figure}
    \centering
    \includegraphics[width=\linewidth]{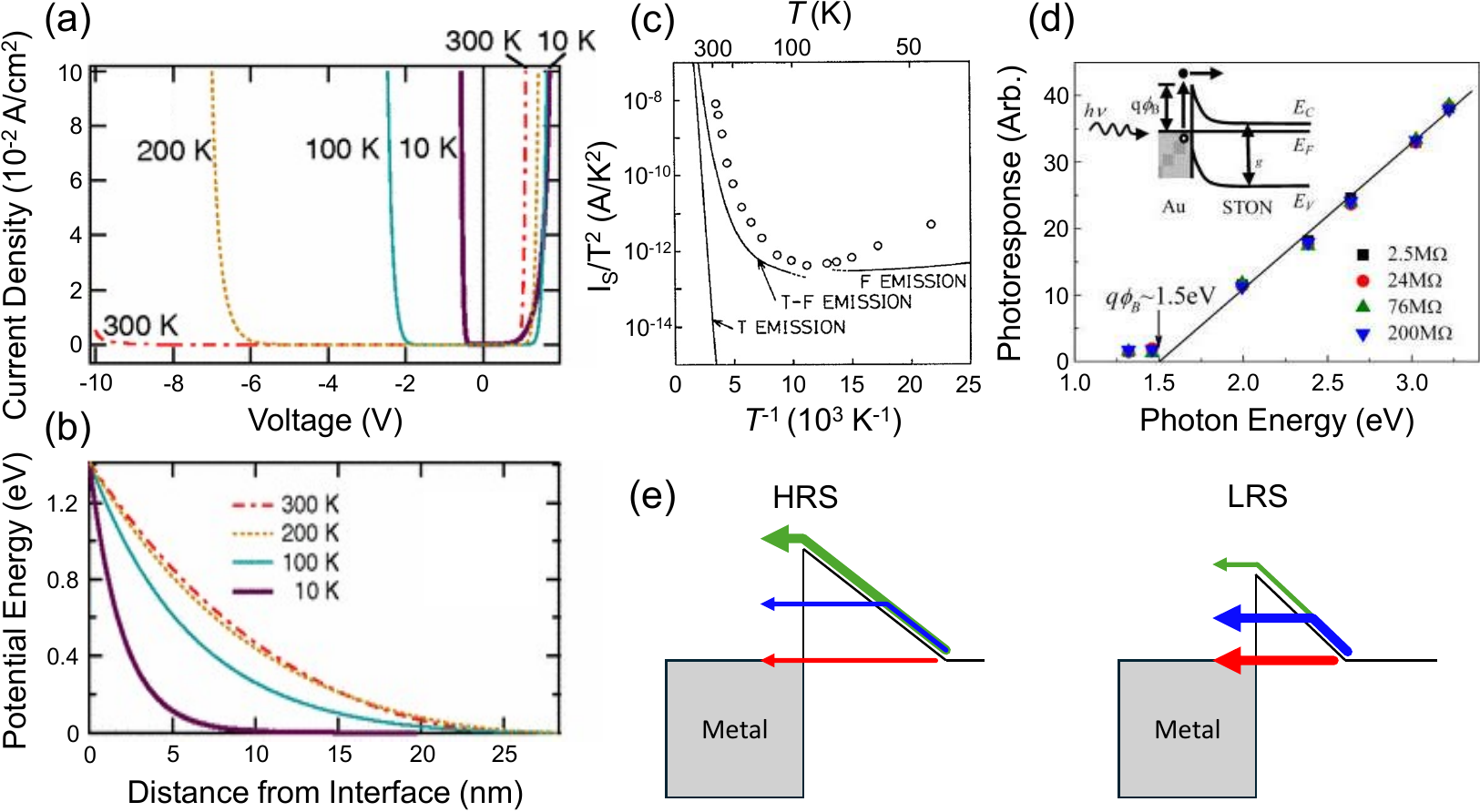}
    \caption{(a) $T$-dependence of Au/Nb:STO I-V curves, demonstrating polarity-reversal at low-$T$. (b) The calculated Schottky barrier energy profile for Au/Nb:STO, assuming a $T$-independent SBH of 1.41~eV.~\cite{Susaki2007PRB} (c) $T$ scaling analysis of the saturation current (I$_\mathrm{S}$), with the expected scaling of thermionic emission, thermionic field emission, and field emission. (d) Resistance state independent SBH measured by internal photoemission, and the inset displays a diagram of the measurement process. (e) Schematic of transport processes, including thermionic emission (green), thermionic field emission (blue), and field emission (red), in the HRS and LRS. A triangular barrier profile is shown as commonly used in the WKB approximation. Figures (a, b) taken from Ref.~\cite{Susaki2007PRB}, (c) taken from Ref.~\cite{Hasegawa1991JAP}, and (d) from Ref.~\cite{Shang2008APL}.}
    \label{fig:24}
\end{figure}

A clear signature of tunneling in M/Nb:STO emerges from $T$-dependent transport, where the thinning of $W_\mathrm{d}$ drives a crossover from TFE to TE at low-$T$ (see Sec.\ref{subsec:IIA1}). The most explicit example of low-$T$ tunneling crossover is the parity reversal observed in Au/, Pt/, and Ni/Nb:STO.~\cite{Susaki2007PRB, Goossens2018JAP,Kim2020APL} In each of these cases, the reverse-bias breakdown voltage decreases with decreasing temperature, leading to a polarity-reversal in rectification at low-$T$ (Fig.~\ref{fig:24}a). This is primarily due to two mechanisms: (1) the lowered potential of Nb:STO in the reverse bias leads to an effective thinning of the $W_\mathrm{d}$ at the Fermi level (Fig.~\ref{fig:13}a) and (2) the decreasing $\epsilon_\mathrm{i}$ at low-$T$ leads to additional $W_\mathrm{d}$ thinning near the interface (Fig.~\ref{fig:24}b).~\cite{Susaki2007PRB} Alternatively, Kim \textit{et al}. diagnoses the transition to low-$T$ tunneling dominance to both SBH reduction and $W_\mathrm{d}$ thinning, where $\epsilon_\mathrm{s}$ increases at low-$T$ but decreases with applied voltage.~\cite{Kim2020APL} This can be understood through the Wentzel-Kramer-Brillouin (WKB) approximation, where the tunneling transmission probability depends exponentially on the barrier width ( i.e., $T \propto \mathrm{exp}(-W_\mathrm{d})$) and on the barrier height (i.e.,  $T \propto \mathrm{exp}(-\phi_\mathrm{B}^{3/2})$) for a triangular barrier.~\cite{Ng2007,Tung1992PRB} Moving on to the forward bias, Hasegawa \textit{et al}. and Susaki \textit{et al}. studied Au/Nb:STO and found anomalous $T$-dependent scaling of the saturation current (I$_\mathrm{S}$) (i.e., the zero voltage condition of the thermionic emission fitting). ~\cite{Susaki2007PRB, Hasegawa1991JAP} Both works conclude that a transition from TFE to FE occurs between 100 and 50~K (Fig.~\ref{fig:24}c).~\cite{Susaki2007PRB, Hasegawa1991JAP} Similar scaling analysis in Ga-In/Nb:STO demonstrated a transition to FE below 130~K.~\cite{Han2004SSI} These works demonstrate that both the SBH and $W_\mathrm{d}$ modulation exponentially impact the tunneling current, and importantly, position TFE as the dominant transport mechanism at room-$T$.

The $T$-dependence is now considered within the framework of the MIS model, which has been well established in M/Nb:STO. Various authors have connected the $T$-dependence of the SBH and $n$ to increased tunneling at low-$T$.~\cite{Cuellar2012PRB, Rana2013PRB, Ruotolo2007PRB,Shimizu1999JAP} 
In the context of the MIS model (see Sec.~\ref{subsec:IIA1}), the observed increase in $n$ and decrease in SBH with decreasing $T$ has been attributed to the voltage-dependent SBH. However, tunneling can also become significant when both $W_\mathrm{d}$ and $\delta$ are small.~\cite{Shimizu1999JAP} In this limit, the intrinsic interfacial layer (i.e., $\delta \approx a_0$) is effectively transparent to electron tunneling.~\cite{Wilt2017ACSAMI, Ng2007,Shimizu1999JAP}
These early studies on M/Nb:STO typically used surface preparation techniques, including chemical etching and/or oxygen annealing,~\cite{Susaki2007PRB, Hasegawa1991JAP,Shimizu1999JAP, Shimizu1997ASS} which indicates such samples possess an intrinsic interfacial layer with minimal RS effects (i.e., $n<1.2$ ). Within the framework of charge trapping/detrapping, this data would more closely reflect the LRS, as the lack of trapping sites does not allow the Schottky barrier buildup necessary for the HRS. Thus, the MIS model is consistent with the observed tunneling behavior and suggests TFE dominates the LRS at room-$T$.

\subsubsection{Trap-Assisted Tunneling}

The presence of OVs and other defect states near the M/Nb:STO interface naturally motivates consideration of TAT as a possible transport pathway and mechanism for RS. It was rationalized by Fujii \textit{et al}. that RS occurs from the creation and destruction of tunneling paths instead of barrier modulation,~\cite{Fujii2007PRB} and Chen \textit{et al}. proposed that resonant tunneling through trapping sites in the LRS explains the CC-C type I-V hysteresis as the traps are available until they are filled by large negative voltage.~\cite{Chen2011APL} Despite these hypotheses, TAT is difficult to experimentally differentiate from TFE and TE, and the electron filling of intermediate states is similar in effect to electron trapping. Fan \textit{et al}. discussed the possibility of TAT and finds a significantly decreased tunneling rate when trapping sites are filled, suggesting that electron traps may play a role in the tunneling behavior in Au/Nb:STO junctions; however, the I-V scaling analysis was consistent with thermionic emission (i.e., $ln(J)$ vs. $V$)~\cite{Fan2017JoMCC} instead of TAT (i.e., $ln(J)$ vs. $V^{-1}$).~\cite{Lim2015E} TAT has also been suggested by Bourim \textit{et al}., where a combination of charge trapping/detrapping and trap-assisted tunneling may occur, and the authors proposed that frequency dependence could separate trapping mechanisms based on characteristic time scales.~\cite{Bourim2014JSSST} Heuristically, charge trapping is expected to exhibit longer trap occupation times, whereas TAT is expected to involve relatively short trap occupation times. While TAT has been used to explain transport in oxide- and nitride-based Schottky junctions,~\cite{Lim2015E,Kim2024TEEM,Al2020MRE, Mahaveer2006JAP} its role in M/Nb:STO is less clear and the lack of scaling evidence suggests a minor contribution.

\subsubsection{Tunneling with Spatial Inhomogeneities }
\label{sec:2B1}

Some experimental observations indicates that RS in M/Nb:STO is not primarily governed by uniform SBH modulation, but may instead involve changes in $W_\mathrm{d}$ and associated spatial inhomogeneities that localize tunneling current in the LRS. Observations by Shang \textit{et al} and Rana \textit{et al}. found that M/Nb:STO junctions with larger Nb doping (i.e., smaller $W_\mathrm{d}$) and enhanced tunneling current exhibit larger on/off ratios.~\cite{Shang2009APL, Rana2012APL} This would indicate that samples with a thin $W_\mathrm{d}$ would be predisposed to tunneling enhancement in the LRS. While capacitance and photoelectric measurements should detect uniform barrier modulation, these measurements found minimal or no change between resistance states, even when the current changes by multiple orders of magnitude (Fig.~\ref{fig:24}d).~\cite{Wang2013APL,Shang2008APL, Lee2011APL} One possible explanation is that tunneling is the dominant mechanism, but constrained to small active areas, which only contributes a small amount to the total contact area and has a stronger influence on the transport.~\cite{Lee2011APL,Kan2013APL, Wang2016ASS} Lee \textit{et al}. proposed that the mean SBH is unchanged between the LRS and HRS, but spatial fluctuations encourage tunneling in the LRS (see Fig.~\ref{fig:13}c and d).~\cite{Lee2011APL} Additional evidence of spatial dependence comes from CAFM and STM studies, suggesting that the macroscopic areal average includes pristine and suppressed Schottky barrier characteristics.~\cite{Wang2016ASS, Roy2013APL,Chen2012JAP} Within the inhomogeneous Schottky barrier picture, RS arises from the evolution of localized tunneling pathways rather than uniform interface modulation, which is consistent with observations that switching is suppressed upon annealing when defect-rich regions are reduced or healed.~\cite{Chen2011APL,Cui2010JAP,Shang2009APL,Wang2016ASS, Wang2013APL} Overall, these works suggest that RS in M/Nb:STO is strongly influenced by both $W_\mathrm{d}$ modulation and spatial inhomogeneities that localize the tunneling current, providing a consistent explanation for the absence of substantial SBH variations in photoelectric and capacitance measurements.

The tunneling-based RS can be understood through a combination barrier inhomogeneities and Schottky barrier modifications, where the LRS is characterized by tunneling and the HRS by thermionic emission (Fig.~\ref{fig:24}e). While some papers suggest either SBH or $W_\mathrm{d}$ modulation,~\cite{Lee2014APLM, Wang2016ASS} the most likely scenario is that both play a role. This has been outlined by Fan \textit{et al}., where the decreased SBH and $W_\mathrm{d}$ in the LRS leads to a dominant tunneling current where thermionic emission fitting are no longer applicable. Even the HRS has some tunneling contributions, despite dominance of thermionic emission.~\cite{Fan2017JoMCC} This makes phenomenological sense, as the $n$ is larger in the LRS than the HRS.~\cite{Mikheev2014NatureCommunications, Fan2017JoMCC, Kunwar2023AEM} A more general transport description therefore requires treating the transport as smoothly evolving within a WKB-based framework. This would provide direct information on the evolution of the Schottky barrier between the HRS and LRS, as well as fitting parameters that can be related to the underlying voltage-dependent RS mechanism.

\subsection{Oxygen Vacancy Mechanisms}

\begin{figure}
    \centering
    \includegraphics[width=\linewidth]{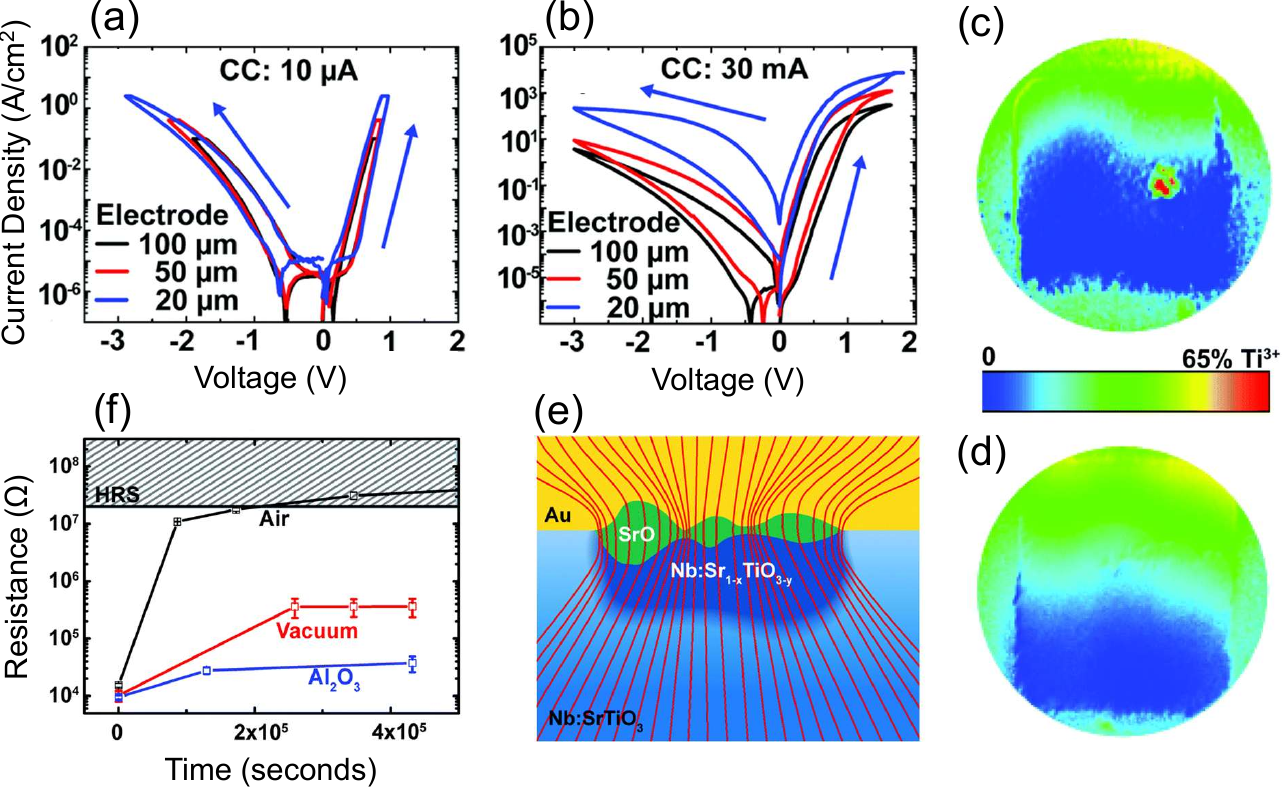}
    \caption{ I-V Curves for Pt/Nb:STO demonstrating (a) area scaling of virgin devices and (b) non-areal scaling of post-forming devices. Spatial dependent Ti L-edge spectroscopy images of Au/Nb:STO after delaminating the Au contact, depicting (c) the presence of a Ti$^{3+}$ hot-spot in the LRS and (d) its absence in the HRS. (e) Cartoon depicting the phase segregation of SrO and Nb:Sr$_\mathrm{1-x}$TiO$_\mathrm{3-y}$, with line density corresponding to current density. (f) Retention of device LRS in air, vacuum, and incorporation of Al$_2$O$_3$ buffer layer. Figures taken from Ref.~\cite{Baeumer2016N}.}
    \label{fig:14}
\end{figure}

OVs are a common type of defect in oxides that act as an $n$-type dopant or charge trap, leading many papers hypothesize that OVs are the dominant RS mechanism in M/Nb:STO.~\cite{Brillson2011JAP,Dharanya2022JNP,Bourim2013CAP,Buzio2012APL,Bian2025FML,Bourim2014JSSST,Chen2011APL,Park2014APL,Lee2014APLM,Quinonez2025APLED,Shen2013APA,Zhong2013CAP,Chen2010APL,Wang2016ASS,Baeumer2016N,Yang2014JAP}
OVs may occupy a doubly charged ($V_\mathrm{O}^{\bullet\bullet}$), singly charged ($V_\mathrm{O}^{\bullet}$), or neutral valence ($V_\mathrm{O}^{\times}$). Due to a typical OV charge of $2+$, it can host two trapped electron states, as demonstrated by the following reaction:~\cite{Kroger1974}
\begin{equation}
    V_O^{\bullet\bullet} + 2e^- \rightleftharpoons V_O^{\times}.
\end{equation}
Using this understanding, various papers have postulated that OVs may be a key trapping centers.~\cite{Brillson2011JAP,Dharanya2022JNP,Bourim2013CAP,Buzio2012APL,Bian2025FML,Bourim2014JSSST,Chen2011APL,Park2014APL,Lee2014APLM,Quinonez2025APLED,Shen2013APA,Zhong2013CAP,Chen2010APL}
In addition to OV-assisted charge trapping/detrapping, VCM-type RS has also been considered. 
Depending on the spatial extent of the active switching region, OV-based RS mechanisms can be broadly classified into two categories: (1) filament-type VCM and (2) interface-type VCM. The interface-type VCM is often forming-free, and the filament-type VCM typically requires electroforming,~\cite{Aussen2023AEM,Dittmann2021AP} Despite this distinction, both mechanisms are governed by field-induced redox processes involving OV migration and redistribution. The physical picture of filament-type VCM based on localized OV-mediated redox reactions is well established.~\cite{Waser2009AM, Joshua2009N, Baeumer2016N, Szot2002PRL, Yang2014JAP, Dittmann2021AP} Dittmann and co-workers  reported Ti$^{3+}$ in the LRS and re-oxided Ti$^{4+}$ in the HRS for filament-type VCM in both M/Nb:STO and M/STO/Nb:STO.~\cite{Baeumer2016N,Cooper2017AM,Funck2020PRB, Baeumer2015NC, Dittmann2021AP,Waser2009AM}
The electroforming process in M/Nb:STO requires thermal activation,~\cite{Baeumer2016N} resulting in a permanently degraded Schottky barrier. This can be achieved at high currents, where Joule heating provides sufficient energy to drive the electroforming process.~\cite{Lenser2014AFM} Yang \textit{et al}. and Baeumer \textit{et al}. have reported on such an electroforming at high bias.~\cite{Yang2014JAP, Baeumer2016N} Additionally, Baeumer \textit{et al}. reported on the dependence of electrode size and compliance current (I$_\mathrm{CC}$) for Pt/Nb:STO devices. When using a I$_\mathrm{CC}$ of 10 $\mu$A for electrodes ranging from 20 to 100 $\mu$m, the devices exhibit I-V curves with area scaling~(Fig.~\ref{fig:14}a), consistent with interface-type RS.~\cite{Baeumer2016N} The I-V profile is also similar with other typical interface-type M/Nb:STO. However, when increasing the I$_\mathrm{CC}$ to 30 mA, the I-V curves no longer scaled with junction area~(Fig.~\ref{fig:14}b), and a soft forming process was observed during the first SET sweep, indicative of CF formation.~\cite{Baeumer2016N} It is interesting that the current density was on the same order (i.e., 10$^3$~A/cm$^2$) for area scaling in the 125~nm diameter (I$_\mathrm{CC}=$1~$\mu$A) and non-area scaling of 100~$\mu$m square electrodes (I$_\mathrm{CC}=$30~mA),~\cite{Baeumer2016N} which seems indicate that the total current is more important than current density and that defect-rich areas will channel the necessary current for Joule heating.
Following this electroforming, both the HRS and LRS will be distinct from the "virgin" device; however, the LRS is not fully ohmic and is still impacted by Schottky physics. The LRS may be dominated by a small fraction of the electrode area, while the HRS will be characterized by the degraded Schottky interface. The degradation of the Schottky barrier can be understood by comparing the onset of exponential conduction in the pre- and post-electroforming I–V curves.~\cite{Yang2014JAP,Baeumer2016N} In summary, the degraded Schottky barrier from high bias electroforming defines the RS through locally concentrated reduction and oxidation processes.

The localization of current in defect-rich areas supplies the necessary current density to permanently degrade the Schottky barrier, inducing a reduced Nb:STO CF and phase-segregated SrO tunnel barrier.~\cite{Baeumer2016N}
Spatially resolved X-ray absorption spectroscopy revealed that filament formation is characterized by an enhanced Ti$^{3+}$ valence state in the LRS (Fig.~\ref{fig:14}c), which is re-oxidized to Ti$^{4+}$ in the HRS (Fig.~\ref{fig:14}d).~\cite{Baeumer2016N} This observation supports the existence of a highly conductive Nb:Sr$_\mathrm{1-x}$TiO$_\mathrm{3-y}$ phase near the interface.~\cite{Baeumer2016N,Szot2002PRL} Additionally, X-ray photoemission spectroscopy revealed phase segregation of SrO at the junction interface (Fig.~\ref{fig:14}e), forming an insulating layer that acts as a tunneling barrier.~\cite{Baeumer2016N} Baeumer \textit{et al.} found that the thinnest SrO layers correspond to the largest current flow, as the thin tunneling barrier minimally impedes conductivity.~\cite{Baeumer2016N} Overall, this highlights that filament conductivity is controlled not only by Ti valence changes but also by the formation and thickness of an interfacial SrO tunneling barrier.


The stability of the redox-based CFs in Nb:STO provides further support for the proposed OV-migration and interfacial oxygen exchange mechanism. Yang \textit{et al}. reported that positive and negative read voltage lead to a two order of magnitude decay of both the HRS and LRS at -0.5~V and 0.5~V, respectively, while at a read voltage of $\pm 0.1~V$, only the LRS showed decay, with the resistance increasing by roughly one order of magnitude over the 250 second measurement window.~\cite{Yang2014JAP} The LRS decay with a small read voltage was attributed to OV diffusion from $V_\mathrm{bi}$.~\cite{Yang2014JAP} The influence of ambient condition was investigated by Baeumer \textit{et al}. by comparing the retention of the LRS between samples stored in air and vacuum. While the samples stored in air decayed to the HRS within $2 \cdot 10^5$~seconds, the vacuum condition only had a one to two order of magnitude decay over $4 \cdot 10^5$~seconds~(Fig.~\ref{fig:14}f).~\cite{Baeumer2016N} This was further improved by insertion of a low oxygen mobility layer of Al$_2$O$_3$ (i.e., Pt/Al$_2$O$_3$/Nb:STO), reducing the decay to less than one order of magnitude over the same time frame.~\cite{Baeumer2016N} Similar retention enhancement was observed in Au/STO/Nb:STO 
hetero-junctions.~\cite{Baeumer2015NC} Together, the work from Yang \textit{et al}. and Baeumer \textit{et al}. support coupled mechanisms of OV migration and interfacial oxygen exchange influencing the stability of filament-type RS.  

The direct evidence to support the interface-type VCM mechanism in M/Nb:STO is rare. It was reported that the current rotation sequence depends on the initial distribution of OVs. Lee \textit{et al}. proposed a theoretical model based on near-interface OV-migration, where OVs migrate between the interface and depletion region instead of between the depletion region and the bulk of Nb:STO.~\cite{Lee2014APLM} In this case, the near-interface model favors CC-C type I-V hysteresis, which is opposite of the conventional C-CC type switching in TiO$_2$. More direct evidence besides energy electron loss spectroscopy to reveal the role and the distribution of OVs will be critical for the confirmation of this mechanism.~\cite{Lee2014APLM, Park2014APL} 

\subsection{Connecting Switching Mechanisms}
\begin{figure}[t]
    \centering
    \includegraphics[width=0.8\linewidth]{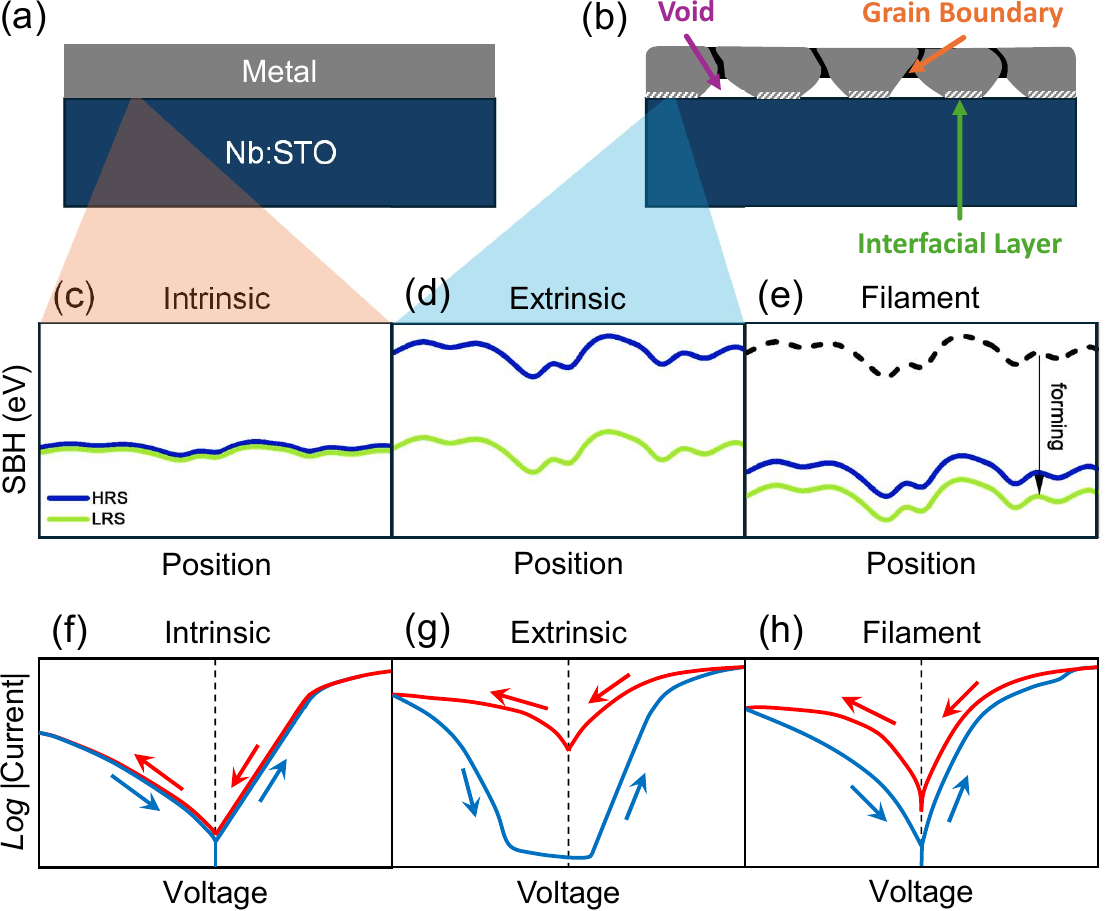}
    \caption{ A cartoon depiction of the M/Nb:STO interface for the (a) intrinsic and (b) extrinsic cases, where the latter involves the formation of an extrinsic interfacial layer in addition to voids and grain boundaries that characterize the electrode microstructure. A schematic depiction of spatial variations of the SBH in (c) the intrinsic case, where the standard deviation and RS is small, and (d) the extrinsic case, where the standard deviation and RS is large. For the extrinsic case, RS occurs in areas with metal-oxide contact. (e) In the event of electroforming, the SBH is suppressed and RS occurs via VCM. Illustrative I-V curves for the (f) intrinsic, (g) extrinsic, and (h) filamentary cases.}
    
    \label{fig:22}
\end{figure}

In a memristor with the structure even as simple as M/Nb:STO, the observed RS behavior is still complicated and the I-V transport can be controlled by a variety of barrier modulation and transport mechanisms depending on Nb:STO interface quality, electrode material/quality, metallization, and applied voltage. In a M/Nb:STO device with an intrinsic interface layer (case 1, Fig.~\ref{fig:22}a and c), I-V curves show rectifying behavior but do not exhibit obvious hysteresis (Fig.~\ref{fig:22}f).~\cite{Mikheev2014NatureCommunications, Buzio2024JoPDAP} In a M/Nb:STO device with a substantial extrinsic interfacial layer (case 2, Fig.~\ref{fig:22}b and d, SBH modulation mechanism), a higher SBH (compared to case 1) is often observed. This results in an extremely insulating  HRS (due to the increased SBH and $W_\mathrm{d}$) and electron trapping in the extrinsic interfacial layer has been widely accepted as the mechanism.~\cite{Fan2017JoMCC, Mikheev2014NatureCommunications, Kunwar2023AEM} The detrapping process under a positive bias can simultaneously lower Schottky interface and enhance the tunneling contribution in the LRS.~\cite{Fan2017JoMCC} However, the SBH modulation mechanism contradicts the limited SBH change between HRS and LRS, reported in photoemission measurements.~\cite{Shang2008APL, Lee2011APL} There is evidence that this SBH modulation might not occur uniformly under the whole electrode. This is because the extrinsic interfacial layer could be highly spatially inhomogeneous, depending on the fabrication process. Thus, it is reasonable that only a (small) portion of the electrode (can be patches under an electrode) effectively goes through the charge trapping/detrapping and tunneling process (Fig.~\ref{fig:22}b).~\cite{Wang2016ASS, Roy2013APL,Chen2012JAP}  If these patches are uniformly distributed under the whole electrode, electrode size scaling can still be valid.~\cite{Wang2013APL, Baeumer2016N} This spatially inhomogeneous picture provides a plausible reconciliation between these two mechanisms. A complete physical picture is following: the interface between M and Nb:STO is extremely inhomogeneous and only a small portion of the top electrode formed a suitable extrinsic interfacial layer which allows trapping/detrapping and SBH modulation. HRS is dominated by Schottky interface and LRS is dominated by tunneling. Both occur at these local areas, rather under the whole electrode area. Therefore, the interface-type RS we defined in M/Nb:STO is often not true under the whole electrode, rather an interfacial effect at some local regions. 

The reduced effective contact area, considering grain boundaries and voids, would give a smaller junction capacitance than expected. As discussed in Sec.~\ref{subsec:IIA1}, the M/Nb:STO junction capacitance is formed by series contributions of $C_\mathrm{i} = \epsilon_\mathrm{i}A/\delta$ and $C_\mathrm{d} = \epsilon_\mathrm{S}A/W_\mathrm{d}$. The metal contact voids would instead give you a term related to the $C_\mathrm{void} = \epsilon_0 A_\mathrm{void}/d_\mathrm{void}$. To the first order the impact of voids can be considered as a parallel contribution:
\begin{equation}
    C_\mathrm{eff} = A[(1-f_v)(\frac{\delta}{\epsilon_\mathrm{i}} + \frac{W_\mathrm{d}}{\epsilon_\mathrm{S}})^{-1} + f_v \frac{\epsilon_0}{d_v}],
\end{equation}
where $A$ is the junction area, $f_v$ is the areal fraction of voids, and $d_v$ is the average depth of void. In the extrinsic limit, where the interfacial capacitance dominates the Schottky response (i.e. $C_\mathrm{i}<<C_\mathrm{d}$), increasing the void fraction reduces the overall effective capacitance. Such a reduced active electrode area may also explain the seemingly counterintuitive observation that electroforming occurs in large-area electrodes at a given nominal current density, but not in smaller-area electrodes at a comparable current density.\cite{Baeumer2016N}  If conduction and forming are localized to only a small fraction of the nominal electrode area, the actual local current density in large-area devices can be substantially higher than the calculated value, causing the effective current density to be underestimated and thereby facilitating electroforming.

In the above physical pictures, no electroforming is involved and therefore filament mechanism is excluded for cases 1 and 2. However, when the applied positive bias exceeds a threshold (case 3, Fig. ~\ref{fig:22}e and h), OV-based CFs will form and dominate switching.~\cite{Baeumer2016N, Yang2014JAP} The formed CF may  be located near some of these patches with relatively lower SBHs. This is why HRS in these cases are more conducting than HRS in case 2 (Fig. ~\ref{fig:22}g and h) as electroforming significantly reduces SBH (Fig. ~\ref{fig:22}e). There are different length scales that lead to such barrier inhomogeneities as shown in Fig. ~\ref{fig:22}c-e. The largest scale is due to the microstructure of the electrode. On the nano- and micro-scale, various crystalline imperfections, such Ti vacancies, Nb segregation, SrO/TiO$_2$ terminations, dislocations, and OV patches, lead to variations in the carrier concentration and bonding environments. This logic follows from certain tunneling hotspots via barrier inhomogeneities, where current funneling and the subsequent Joule heating leads to filament formation in high defect areas under the electrode (Fig.~\ref{fig:14}c).

\section{Factors Influencing Resistive Switching}
\subsection{Electrode Materials}
To probe M/Nb:STO Schottky junctions, forming a reliable ohmic contact to the back side of Nb:STO is essential. High-work-function metals (e.g., Pt, Au, Ni) typically form rectifying Schottky junctions on Nb:STO, whereas low-work-function metals (e.g., In, Al, Ti) produce ohmic behavior. While the Schottky-Mott rule links the metal work function directly to the SBH, interface preparation and surface quality strongly influence the effective barrier and transport. For example, In press contacts require surface roughness or melting to achieve ohmic behavior, whereas Ti and Al form robust ohmic contacts via standard deposition techniques.~\cite{Cui2007MSEB} Additionally, the thermodynamics of metallization plays a role, where the nucleation density and surface energy impact bonding between the metal and Nb:STO.~\cite{Mattox1973TSF, Ng2007,Li2023APL} 
For instance, large Au contacts can give ohmic behavior through PLD deposition or by abrading the Nb:STO surface before sputter deposition.~\cite{Chen2011APL, Wang2016APL}.
Generally among high work-function metals, experimentally extracted SBHs are often lower than the Schottky–Mott rule~(Table~\ref{tab:1}),  reflecting non-ideal interfaces and parallel transport paths.

\begin{table}[t]
\caption{SBH measurements on various M and CO contacts. IPE: Internal Photoemission, HXPES: Hard X-ray Photoemission Spectroscopy, PES: Photoemission Spectroscopy, PER: Photoelectric Response.}
\centering
\begin{tabular}{ccccc}
\hline\hline
 & SM (eV) & I-V (eV) & C-V (V) & Photoelectric (eV)\\ 
\hline
Au &  1.5~\cite{Michaelson1977JAP}  & 0.83, $n$:1.60 (0.01 wt.\%)~\cite{Buzio2024JoPDAP} & 1.77 (0.01 wt.\%)~\cite{Shimizu1999JAP} &  1.47 (0.01 wt.\%), IPE~\cite{Hikita2011APL}\\
&  & 0.88, $n$:2.16 (0.1 wt.\%)~\cite{Park2008JAP} &  0.98 (0.1 wt.\%)~\cite{Park2008JAP} &  0.6 (0.05 wt.\%), HXPES~\cite{Ohsawa2021JoPCC}\\
&  & 1.60, $n$:1.71 (1 wt.\%)~\cite{Cui2007MSEB} &  &  \\
\hline
Pt &  1.7~\cite{Michaelson1977JAP}  &  0.75, $n$:1.10 (0.01 wt.\%)~\cite{Buzio2024JoPDAP} &  1.17 (0.1 wt.\%)~\cite{Park2008JAP}& 0.75 (0.05 wt.\%), HXPES~\cite{Ohashi2012APL}\\
& &  0.77, $n$:1.60 (0.1 wt.\%)~\cite{Park2008JAP} &   &0.8 (0.6 wt.\%), HXPES~\cite{Hirose2015APL}\\
\hline
SRO &  1.1~\cite{Minohara2007APL} &  1.17, $n$:1.4(0.01 wt.\%)~\cite{Fujii2007PRB}&  1.50 (0.01 wt.\%)~\cite{Fujii2007PRB}& 1.47 (0.01 wt.\%), IPE~\cite{Hikita2007APL} \\
&  & 0.70, $n$:1.81 (0.5 wt.\%)~\cite{Hikita2007APL} & 0.90 (0.5 wt.\%)~\cite{Fujii2007PRB} &  1.20 (0.05 wt.\%), PES~\cite{Minohara2007APL} \\
&  & 0.58, $n$:2.2 (1 wt.\%)~\cite{Fujii2007PRB} &  & 1.31 (0.5 wt.\%), IPE~\cite{Hikita2007APL}  \\
\hline
LSMO &  0.7~\cite{Minohara2007APL} &  0.95, $n$:1.08 (0.01 wt.\%)~\cite{Postma2004JAP} & 0.50 (0.01 wt.\%)~\cite{Sawa2005APL} & 1.20 (0.05 wt.\%), PES~\cite{Minohara2007APL} \\
& & 0.65, $n$:1.18 (0.1 wt.\%)~\cite{Postma2004JAP} & 0.75 (1 wt.\%)~\cite{Ruotolo2007PRB} & \\
\hline
YBCO &  2.1~\cite{Hao2016APL} &  0.96 (0.7 wt.\%)~\cite{Hao2016APL} & 0.74 (0.5 wt.\%)~\cite{Yoshida1991JournalofAppliedPhysics} & 2.0 (0.7 wt.\%), PER~\cite{Hao2016APL}\\
& & & 1.6 (0.7 wt.\%)~\cite{Hao2016APL}& \\
\hline
BKBO &  &  1.67, $n$:1.19 (0.01 wt.\%)~\cite{Yamamoto1998JJAP} & 1.73 (0.01 wt.\%)~\cite{Suzuki1997JAP}  &  \\
&  &  1.60, $n$:1.45 (0.05 wt.\%)~\cite{Yamamoto1998JJAP} &  &  \\
&  &  1.40, $n$:1.79 (0.5 wt.\%)~\cite{Yamamoto1998JJAP} &  &  \\
\hline\hline
\label{tab:1}
\end{tabular}
\end{table}

\begin{figure}
    \centering
    \includegraphics[width=0.5\linewidth]{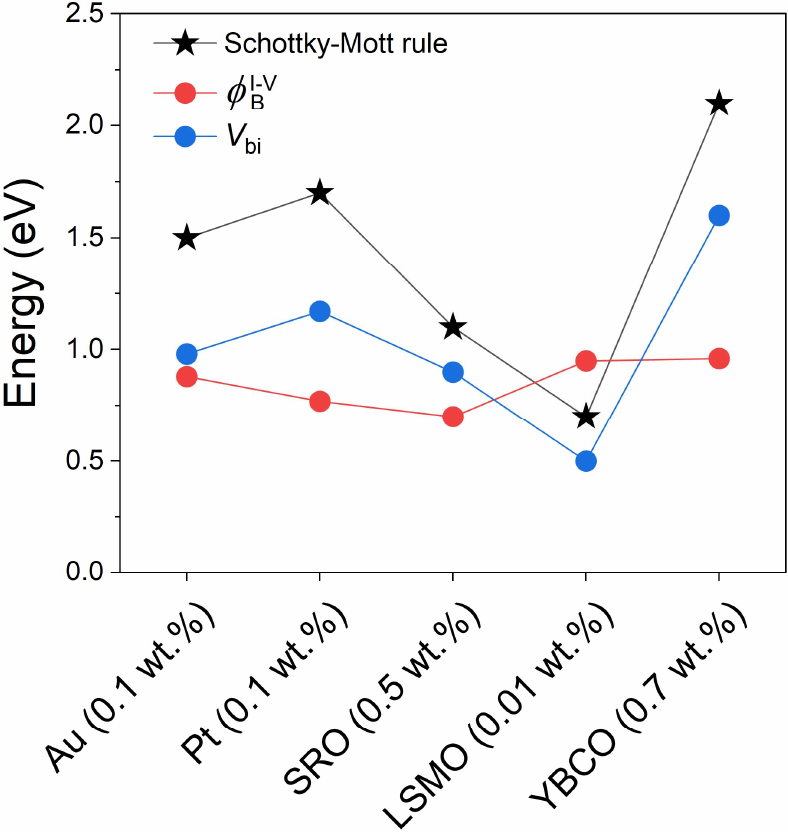}     
    \caption{Comparison of I-V and C-V measurements of the SBH, compared with the Schottky-Mott rule. See Table~\ref{tab:1} for references.}
    \label{fig:15}
\end{figure}

Conductive oxides (COs), such as SrRuO$_3$ (SRO), La$_\mathrm{1-x}$Sr$_\mathrm{x}$MnO$_3$ (LSMO), Ba$_\mathrm{1-x}$K$_\mathrm{x}$BiO$_3$ (BKBO), and  YBa$_2$Cu$_3$O$_\mathrm{7-x}$ (YBCO), provide a route towards RS in epitaxial Schottky junctions on Nb:STO, revealing many similarities with noble metal contacts. The overlap between M/ and CO/Nb:STO, such as increased tunneling at low-$T$ or high Nb doping,~\cite{Hikita2008PRB,Sawa2005APL,Kurij2016TSF,Ruotolo2007PRB,Postma2004JAP, Rana2013PRB, Fujii2007PRB} reflects changes in the depletion region of Nb:STO.~\cite{Kim2020APL} This has a strong influence on Schottky barrier transport, which can be seen in the divergence of Schottky barrier measurements in I-V and C-V measurements, as shown in Fig.~\ref{fig:15}. Because C-V probes $W_\mathrm{d}$ and $V_\mathrm{bi}$, whereas I-V reflects carrier transport across the barrier, SBHs obtained from C-V generally correlate more closely with the Schottky-Mott rule. In contrast, SBHs derived from I-V measurements are often underestimated due to non-ideal transport, originating from spatial inhomogeneities and tunneling pathways. These trends hold true for Au, Pt, SRO, and YBCO (Fig.~\ref{fig:15}); however, LSMO is a notable outlier: a polar discontinuity at the interface leads to an underestimated SBH from C-V analysis and overestimated SBH from I-V compared to the Schottky-Mott rule.~\cite{Minohara2007APL, Hikita2009PRB, Minohara2010PRB, Minohara2012PRB} Therefore, these trends unify observations across M- and CO-based junctions, underscoring that transport and RS in Nb:STO Schottky devices are governed primarily by interfacial structure rather than metal work function alone. This can be seen in Tab.~\ref{tab:1}, where higher Nb-doping suppresses the SBH in epitaxial SRO, LSMO, and BKBO, but Pt and Au do no exhibit such a trend and depends more on deposition method.

The current rotation direction in I-V hysteresis loops of M/Nb:STO systems is CC-C. Both CF filament mechanism and trap/detrapping mechanism can contribute to the CC-C current rotation sequence. As discussed before, both mechanisms are possible in M/Nb:STO junctions and they have been extensively discussed above. In CO/Nb:STO, the current rotation direction of I-V hysteresis loops are actually the same as that in M/Nb:STO systems. It should be noted that most of CO/Nb:STO papers defined their current opposite to M/Nb:STO systems which makes them look having opposite current rotation directions.\cite{Fujii2007PRB,Fujii2005APL, Zhang2009APL, Jia2025MSEB} The importance of OV in CO/Nb:STO systems is confirmed by oxygen pressure dependent study~\cite{Wang2019CPB} and oxygen annealing studies.\cite{Sawa2008MaterialsToday,Song2026MSSP}  Since most of the I-V hysteresis loops in CO/Nb:STO systems is CC-C, indicating OVs are more tied to charge trapping/detrapping and/or CFs.~\cite{Song2026MSSP} It should be noted that OV drift can modulate barrier height and/or width and therefore change I-V behavior. This has been widely discussed in TiO$_x$,~\cite{Yang2008NN} SrFeO$_{3-\delta}$,~\cite{Su2024AMI} and Pr$_{0.7}$Ca$_{0.3}$MnO$_3$.~\cite{Baek2017N} However, such an OV drift mechanism often results in C-CC type current rotation sequence. Jia \textit{et al.} reported that PLD-deposited YBCO/Nb:STO devices show this C-CC type switching with relatively small positive voltage while the same device can turn into a CC-C type switching with a higher positive bias.\cite{Jia2025MSEB}  It is likely that with a smaller bias, the device switches by OV redistribution modulated Schottky barrier. Once the bias is over a threshold, a equivalent forming occurs and therefore the switching is then dominated by OV-based CF, where the device switches to a CC-C current rotation sequence. Such an observation of changing current rotation sequence is common. In a VO$_2$/LSMO bilayer, volatile switching and non-volatile switching can be tuned by changing the magnitude of the positive bias, which confirming the location of OVs can determine the switching occurring in either VO$_2$ or LSMO.~\cite{Kunwar2024AMI} In fact, such a polarity change has been often reported in systems based on OV migration via tuning voltage magnitude including Pt/TiO$_{2-x}$/Pt~\cite{Yang2008NN} and Au/Sr$_2$TiO$_{4-x}$/Nb:STO.~\cite{Shibuya2010AM}  


\subsection{Impact of Surface Quality}

The methods used for electrode deposition (e.g., evaporation, sputtering, PLD) and surface treatments of substrates (e.g., annealing, etching) significantly impact the interface structure, defect density, and the presence of extrinsic interfacial layers. These factors are critical in determining the presence and characteristics of RS. This section focuses on how to tune interface quality as well as RS.

\subsubsection{Metallization Methods}
\begin{figure}
    \centering
    \includegraphics[width=0.8\linewidth]{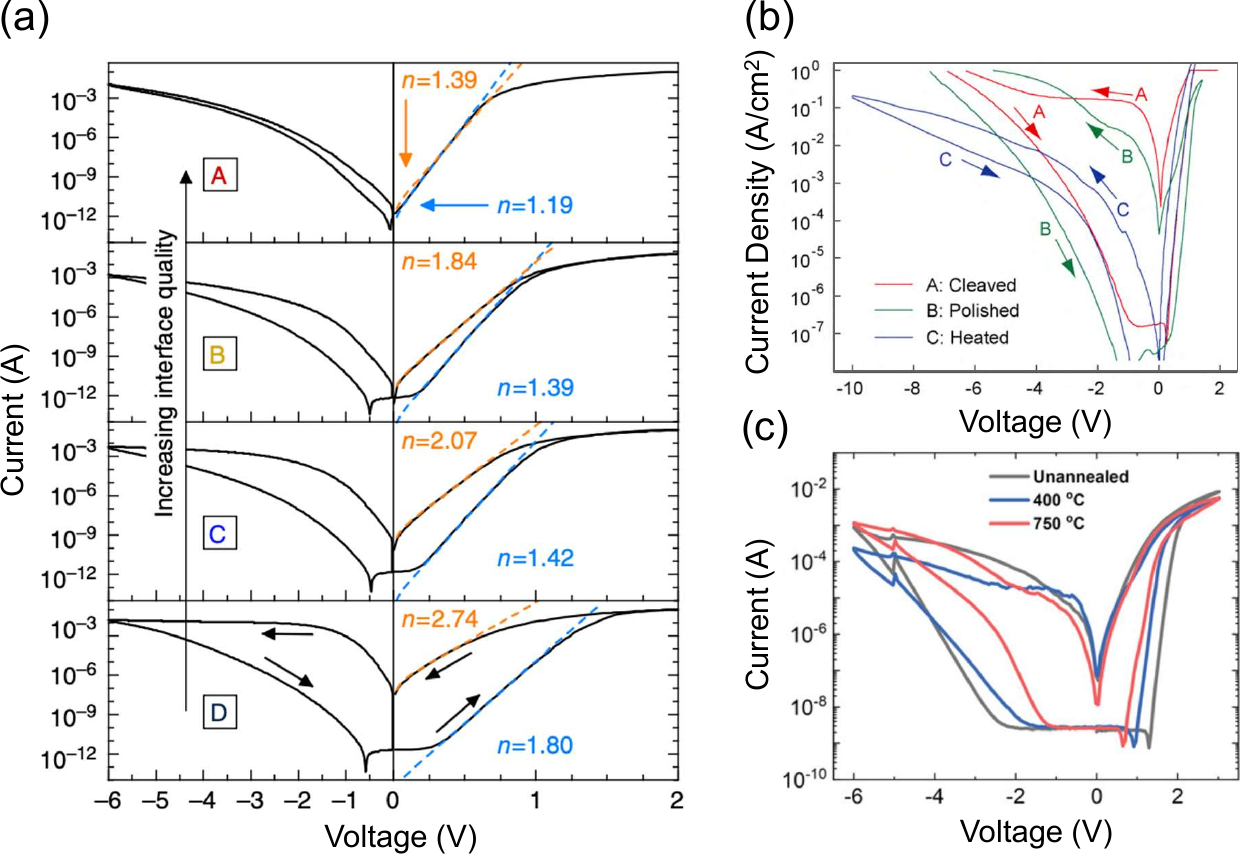}
    \caption{(a) I-V hysteresis and extracted $n$ for different deposition conditions of Pt on Nb:STO. The description of A, B, C, and D can be found in Fig.~\ref{fig:8} caption and Ref.~\cite{Mikheev2014NatureCommunications}. (b) Comparison of I-V hysteresis in Pt/Nb:STO junction formed on cleaved, polished, and heated Nb:STO. (c) Comparison of I-V Hysteresis in Au/Nb:STO by annealing as-delivered substrates at 400 and 750 $^\circ$C for 2 hours. Figure (a) taken from Ref.~\cite{Mikheev2014NatureCommunications}, (b) from Ref.~\cite{Kunwar2023AEM}, and (c) from Ref.~\cite{Li2010MSEB}.}
    \label{fig:16}
\end{figure}

The deposition kinetics play a significant role in metal-oxide bonding. Mikheev \textit{et al}. reported that the deposition conditions, and consequently the interface quality, plays a critical role in determining the RS hysteresis by changing the process conditions for electrode deposition.~\cite{Mikheev2014NatureCommunications} Large I-V hysteresis is observed in samples when top Pt electrodes are deposited by standard e-beam evaporation at room temperature (i.e., sample D in Fig.~\ref{fig:16}). I-V hysteresis loop becomes suppressed when Pt electrode crystal quality is improved. When the top electrode Pt is epitaxially deposited at 825 $^\circ$C via DC sputtering, I-V hysteresis is nearly absent (i.e., sample A in Fig.~\ref{fig:16}). Since all samples were post-Pt deposition annealed at 800 $^\circ$C in flowing O$_2$ to suppress contributions from OVs,~\cite{Mikheev2014NatureCommunications} Mikheev \textit{et al}. claimed the RS is related to the  interfacial layer rather than OVs. In summary, Mikheev's results clearly showed the better the interface quality, the smaller the ON/OFF ratio.

In addition, M/Nb:STO devices with metal electrodes grown by room temperature PLD also show suppressed RS.~\cite{Buzio2024JoPDAP} While these metal electrodes are not epitaxial, intimate metal-oxide contact is achieved by the plasma bombardment of metal onto the Nb:STO surface. For Au/Nb:STO, the SBH is approximately 0.7-0.8~eV,~\cite{Buzio2024JoPDAP} significantly below the Schottky-Mott rule of 1.5 eV.~\cite{Michaelson1977JAP} 
More strikingly, such junctions are much less sensitive to moisture and the aging effect is substantially reduced for junctions with the PLD-grown Au and Pt.~\cite{Buzio2024JoPDAP} It is interesting to note that both epitaxial Pt electrodes grown at 825 $^\circ$C by DC sputtering~\cite{Mikheev2014NatureCommunications} and Pt electrodes grown at room temperature by PLD~\cite{Buzio2024JoPDAP} on thermally treated Nb:STO show limited RS behavior with the $n$ close to 1.1 and 1.2, respectively. This further emphasizes that $n$ is a key parameter for both the interface quality and RS performance. 

At high growth temperatures, DC sputtering is expected to produce epitaxial Pt electrodes with direct contact on Nb:STO substrates without any physical gap. For PLD growth at room temperature, although these metal electrodes are not epitaxial, the UV-radiation of the plasma plume and the energetic impinging metal species promote intimate metal contact with Nb:STO. Therefore, we speculate that the formation of the interfacial layer is directly correlated with the effectiveness of the physical contact between metal and Nb:STO. When the metal film directly contacts Nb:STO without significant amount of defects (e.g., gap, void, vacancies, \textit{etc}.), the extrinsic interfacial layer is unlikely to form and moisture is unable or difficult to creep into the interface. Therefore, both epitaxial Pt electrodes by sputtering and non-epitaxial Pt electrodes by PLD promote intimate metal-oxide contact to eliminate RS. 

\subsubsection{Surface Treatments for Nb:STO Substrates}

The previous section confirmed the top metal electrode film quality and its contact with Nb:STO plays a critical role in determining the RS and junction properties. Next, let’s discuss how the treatment of Nb:STO substrates affects the RS properties. Annealing of Nb:STO substrates significantly improves surface quality and has typically utilized the "Arkansas" method (i.e., a combination of wet-etching and high-$T$ O$_2$ annealing) to reduce step edge population and promote the TiO$_2$ surface termination.~\cite{Kareev2008APL} This is commonly used for STM measurements, where atomically sharp surfaces are required.~\cite{Buzio2018APL} However, it is typical for macroscopic electrode measurements to pre-anneal Nb:STO substrates for 1-2 hour at 800-1000~$^\circ$C while flowing O$_2$ to fill OVs and improve surface uniformity.~\cite{Buzio2024JoPDAP,Buzio2012APL,Gerbi2014AMI,Park2008JAP, Wang2016ASS}  Further, annealing substrates before growth removes dangling bonds, contaminants, and OVs thereby yielding cleaner M/Nb:STO junctions; however, over annealing can results in unwanted surface reconstruction and procedures should be optimized.~\cite{Son2024JKPS, Kunwar2023AEM} This directly affects the RS hysteresis and ON/OFF.~\cite{Gerbi2014AMI, Kunwar2023AEM, Li2010MSEB, Chen2016SAM,Li2023APL} 

We can now examine how these annealing protocol impact the RS performance of Nb:STO junctions. For example, Li \textit{et al}. found that I-V hysteresis is suppressed if the surface of Nb:STO is thermally treated.~\cite{Li2010MSEB} Comparing as-polished, thermally-treated, and cleaved Nb:STO substrates, it was found that cleaved substrates exhibited the largest I-V hysteresis and thermally-treated substrates exhibited the smallest in Pt/Nb:STO junctions (see Fig.~\ref{fig:16}b).~\cite{Li2010MSEB} In another report, Kunwar \textit{et al}. compared the RS effect of annealing temperature of Nb:STO in air.~\cite{Kunwar2023AEM}  It was reported that I-V hysteresis loops consistently shrink with increasing the Nb:STO annealing temperature with the unannealed Nb:STO sample exhibiting the largest RS hysteresis (see Fig.~\ref{fig:16}c).~\cite{Kunwar2023AEM} These results here and above highlight the RS in M/Nb:STO is directly controlled by the interface quality which includes both the quality of the top electrode and the quality of the substrate surface.

A recent work investigated the influence of the interface contact condition on the RS. Li \textit{et al}. compared two Au/Nb:STO devices with different surface treatments.~\cite{Li2023APL} One Nb:STO substrate surface was cleaned \textit{in-situ} (i.e., vacuum annealing to remove surface contamination) followed by metallization and the other Nb:STO substrate was used without surface cleaning.~\cite{Li2023APL} It was found that the junction with \textit{in-situ} surface cleaning didn’t show RS while the sample without surface treatment exhibit regular RS, ~\cite{Li2023APL} which is consistent with discussion above. Strikingly, SEM results show that an intimate contact between metal and the \textit{in-situ} cleaned substrate was formed.~\cite{Li2023APL} In contrast, the untreated Au/Nb:STO interface exhibits poor contact with local physical gaps between Au and Nb:STO substrates.~\cite{Li2023APL} It is likely that surface contamination results in poor electrode contact, contributing to enhanced RS behavior. 

\subsection{Impact of Interface Chemistry}

\subsubsection{Nb Doping Concentration}
\label{sec:Nb doping}

The Nb concentration in Nb:STO plays a critical role in surface crystallinity, treatment effectiveness, $W_\mathrm{D}$, and $\epsilon_\mathrm{s}$, as well as the overall magnitude of the switching response. The most straightforward impact of Nb doping is control over $W_\mathrm{d}$, since $W_\mathrm{d} \propto N_\mathrm{D}^{-1/2}$. Increasing $N_\mathrm{D}$ in Nb:STO was shown to reduce the SBH and increase the $n$ extracted from I–V measurements, in addition to decreasing both $V_\mathrm{bi}$ and $W_\mathrm{d}$ from C–V measurements.~\cite{Fujii2007PRB, Shimizu1999JAP} The SBH lowering due to electron doping (i.e., Nb and OV) has been supported by density functional theory calculations.~\cite{Funck2019AIPA} While some reports support that increasing Nb content leads to enhanced RS,~\cite{Shang2009APL, Rana2012APL} there is an optimal $N_\mathrm{D}$ for RS performance. This has been shown to be a relatively low value of 0.05 wt.\% for SRO/Nb:STO junctions,~\cite{Fujii2007PRB} while for (Pt, Ag)/Nb:STO junctions 0.7 wt.\% gives a 100 times larger ON/OFF ratio than 0.05 wt.\%.~\cite{Li2019PSSA, Chen2011APL} Chen \textit{et al}. further increased $N_\mathrm{D}$ by vacuum annealing at 800 $^\circ$C for 20 hours to enhance the OV concentration, leading to the suppression of RS for both 0.7 wt.\% and 0.05 wt.\% .~\cite{Chen2011APL} These results indicate that RS in M/Nb:STO is strongly dependent on the Nb concentration, with optimal ON/OFF ratios observed near 0.7 wt.\% for noble-metal electrodes.

A secondary impact of Nb doping comes from the introduction of spatial inhomogeneity, which may impact annealing protocols to obtain atomically flat interfaces. A STM study conducted by Buzio \textit{et al}. used ballistic electron emission microscopy (BEEM) to evaluate the spatial dependence of the SBH, finding that an increased Nb content lead to a larger standard deviation in the SBH.~\cite{Buzio2018APL} Additionally, Chen \textit{et al}. used electron beam-induced current (EBIC) and TEM to reveal a high density of dislocation arrays in 0.5 wt.\% Nb:STO, which may be related to the enhanced RS at high doping.~\cite{Chen2016SAM} These spatial inhomogeneities from Nb-doping may be exacerbated by oxygen annealing, as Postma \textit{et al}. reported via AFM measurements that 1 bar of oxygen at 950~$^\circ$C leads to an increased Nb concentration at the surface for 0.1 wt.\% Nb:STO.~\cite{Postma2004JAP} However, Marshall \textit{et al}. found an absence of such Nb segregation up to 1500~$^\circ$C in ultra-high vacuum for 0.7 wt.\% Nb:STO.~\cite{Marshall2011PRB} While the annealing effects of Nb doping are less certain, enhanced doping levels lead to spatial inhomogeneity which may be linked to the enhanced RS at moderate doping levels.

\subsubsection{Aging and Environment}

\begin{figure}
    \centering
    \includegraphics[width=\linewidth]{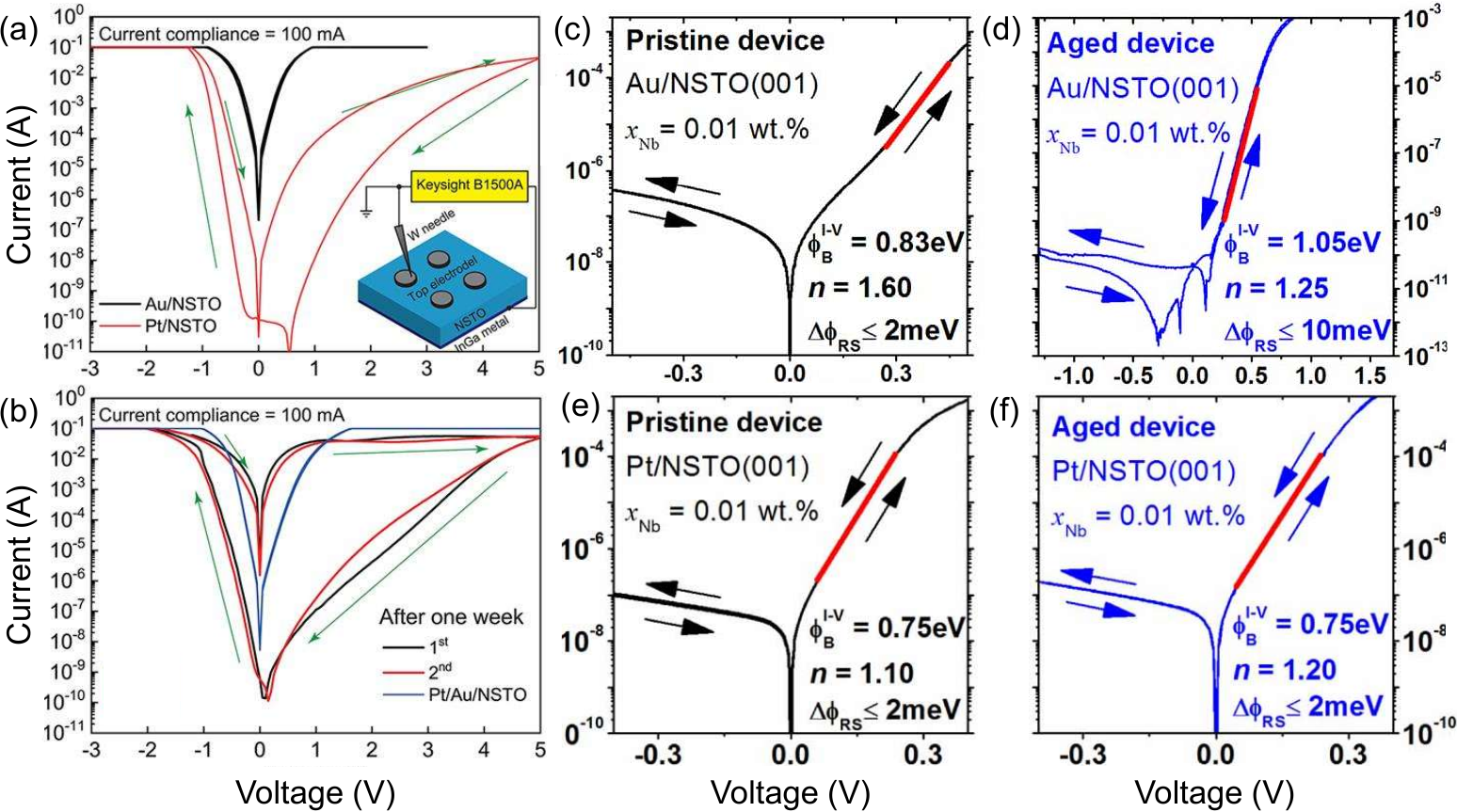}
    \caption{(a) As-grown I-V curves of Au/ and Pt/Nb:STO. (b) Enhancement of RS in Au/Nb:STO observed after one week, and unchanged RS in Pt-capped Au/Nb:STO. The voltage polarity of (a) and (b) is opposite to what is defined in Figs. 5 and 6. Comparison of PLD-deposited Au/Nb:STO in the (c) pristine device and (d) after 60 months in air. (e, f) The same comparison as (c, d) for PLD-deposited Pt/Nb:STO. Figures (a, b) taken from Ref.~\cite{Hirose2019JAP} and (c-f) taken from Ref.~\cite{Buzio2024JoPDAP}.}
    \label{fig:aging}
\end{figure}

Aging effects in M/Nb:STO junctions are largely governed by dynamic interfacial defect chemistry, where the interaction with the ambient environment progressively alter the RS characteristics over time. An initial work by Hirose \textit{et al}. showed that aging in air strengthened RS behavior and increased the SBH in sputtered Au/Nb:STO devices~(Fig.~\ref{fig:aging}a and b).~\cite{Hirose2019JAP} Ohsawa \textit{et al}. similarly found that sputtered Au/Nb:STO junctions exhibited RS only after 12 days in air, indicating that the active interface forms progressively after fabrication.~\cite{Ohsawa2021JoPCC} To understand which gas species in air control this aging effect and dominate the RS, a more recent study compared I-V curves of e-beam evaporated Au/Nb:STO devices in different environments. RS in these devices collapsed when heated in dry O$_2$ or dry N$_2$, while the RS can be restored after exposure to air or humid N$_2$ (Fig.~\ref{fig:10}j), whereas no comparable changes were observed in dry O$_2$ or dry N$_2$.~\cite{Kunwar2023AEM} This result concludes humidity/moisture plays a key role in enabling RS in Au/Nb:STO.

Therefore, it is surprising that the aging effect can be mitigated by minimizing the permeability of the electrode. For instance, Hirose \textit{et al}. found that capping Au/Nb:STO junctions with Pt substantially reduced the aging effect~(Fig.~\ref{fig:aging}a and b).~\cite{Hirose2019JAP} Another route to suppress aging has been shown through intimate metal-oxide contact with PLD deposited contacts. Buzio \textit{et al}. reported almost no aging effect in PLD-grown Au/ and Pt/Nb:STO junctions over 60 months, suggesting that intimate metal-oxide contacts can be inherently stable.~\cite{Buzio2024JoPDAP} Among these devices, Au/Nb:STO exhibited more pronounced aging, including moderate reverse-bias RS, a shift of the minimum current away from zero bias, and a decreased ideality factor $n$ (Fig.~\ref{fig:aging}c and d), whereas Pt/Nb:STO remained largely unchanged (Fig.~\ref{fig:aging}e and f). These studies indicate that electrode permeability and imperfect contact between electrode and Nb:STO facilitate the long-term chemical evolution responsible for aging in M/Nb:STO devices. 

RS in M/Nb:STO can also be sensitive to the oxygen content. When comparing measurements in air and vacuum, (Au, Pt)/Nb:STO have shown a more conductive LRS.~\cite{Wang2016ASS, Bourim2014JSSST, Buzio2012APL, Bourim2013CAP}  Buzio \textit{ et al}. found that increasing the oxygen partial pressure lead to decreased reverse bias leakage current.~\cite{Buzio2012APL} This suggests that more interfacial oxidation leads to a more insulating Schottky junction. Both Buzio \textit{et al}. and Bian \textit{et al}. argue that  the absorption or desorption of oxygen at the interface by varying the oxygen partial pressure plays an important role in RS.~\cite{Buzio2012APL, Bian2025FML}

\subsection{Impact of Voltage}
\label{Sec:impactofvoltage}

\begin{figure}
    \centering
    \includegraphics[width=\linewidth]{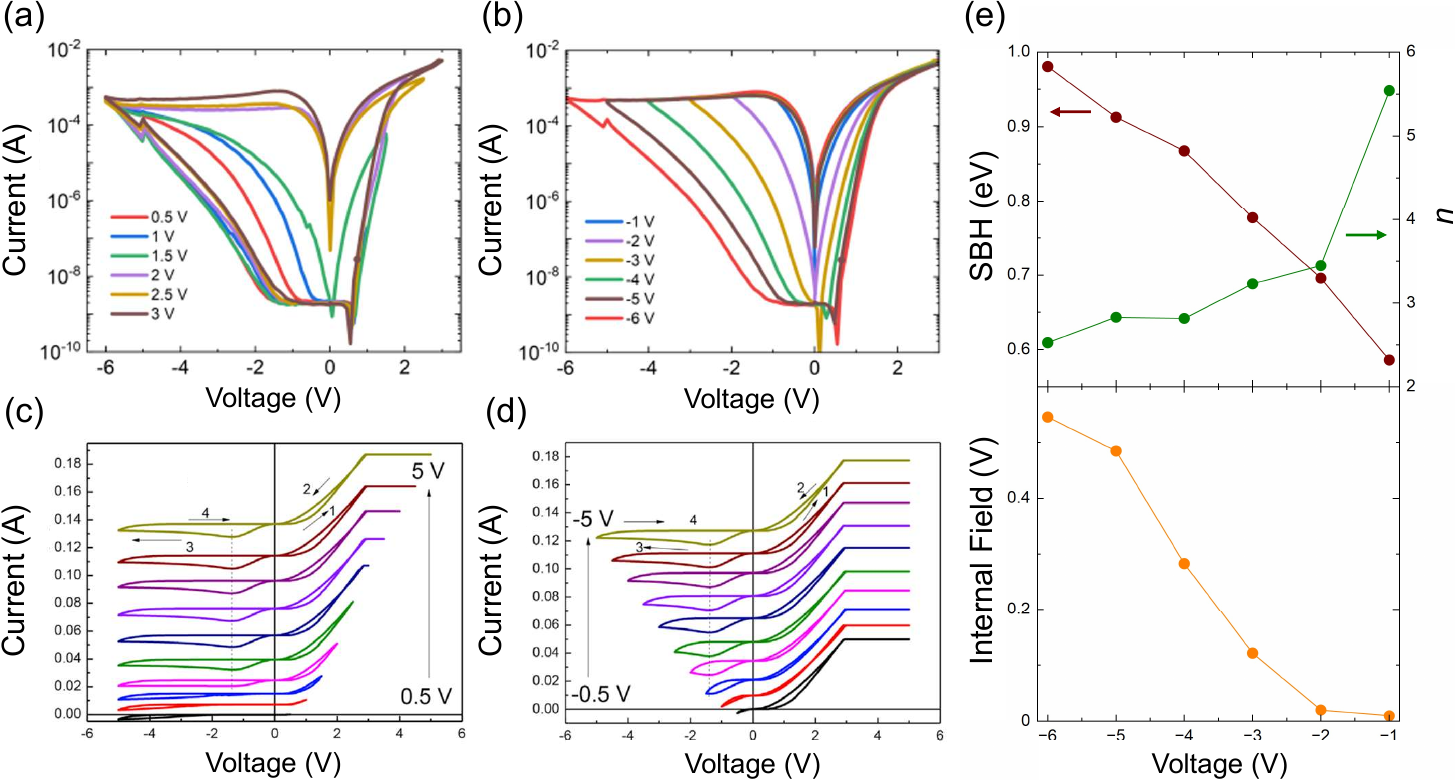}
    \caption{Voltage dependence of e-beam evaporated Au/Nb:STO plotted in semi-log scale from (a) 0.5 to 3~$V$in the positive bias and (b) - 1 to 6~$V$in the negative bias. Voltage dependence of sputtered Au/Nb:STO plotted in linear scale from (c) 0.5 to 5~$V$in the positive bias, and -0.5 to -5~$V$in the negative bias. (e) Negative bias voltage dependence of SBH, $n$, and the internal field (i.e, global minimum in current) extracted from (b). Figures (a) and (b) taken from Ref.~\cite{Kunwar2023AEM} and (c) and (d) from Ref.~\cite{Li2018PSSA}.}
    \label{fig:17}
\end{figure}

The applied voltage plays a critical role in controlling the RS behavior and the shape of the I-V hysteresis loop. Negative differential resistance (NDR) is commonly observed in the negative-bias regime of the LRS after the junction has undergone a complete detrapping process induced by a large positive bias (i.e., SET operation), as seen in Fig.~\ref{fig:17}a-d.~\cite{Wang2018NRL, Li2018PSSA, Kunwar2023AEM, Goossens2018JAP} At small negative bias, the current initially increases because electron transport is dominated by tunneling through the interfacial layer, while the effective SBH and $W_\mathrm{d}$ remain relatively small. As the negative bias is further increased, electrons are progressively trapped near the interface, leading to an increase in the effective SBH and a widening of $W_\mathrm{d}$. When the suppression of tunneling caused by trap-induced barrier modification exceeds the enhancement due to the increasing electric field, the current begins to decrease with increasing bias, resulting in the emergence of NDR.~\cite{Goossens2018JAP, Kunwar2023AEM} Thus, NDR in M/Nb:STO serves as a strong indicator of the effectiveness of interfacial trapping and detrapping processes.


The effectiveness of the detrapping process is correlated with the opening of the I-V hysteresis loop. 
Often, a positive 3~$V$ is necessary to achieve a significant detrapping which restores the LRS. As seen in Fig.~\ref{fig:17}a and c, 2~$V$ is not enough to complete the detrapping, preventing full recovery of the LRS.~\cite{Li2018PSSA,Kunwar2023AEM} Consequently, optimization of RS requires an appropriate balance between positive and negative bias conditions. While Fan \textit{et al}. has simulated I-V behavior within the MIS model incorporating charge trapping/detrapping ,~\cite{Fan2017JoMCC}, the development of a more microscopic transport model that explicitly considers trap energy levels and trap concentrations would be valuable, analogous to models developed for OV migration.~\cite{Lee2014APLM,Sung2013APL,Quinonez2025APLED} 

To summarize the impact of voltage on Nb:STO, the range and polarity of the voltage bias determines the overall switching mechanisms (e.g., interface-type, filamentary-type) and insulating nature of the Schottky diode. In the negative voltage (reverse bias) regime, a large voltage promotes charge trapping, building the SBH and $W_\mathrm{d}$~(Fig.~\ref{fig:17}e).~\cite{Goossens2018JAP, Kunwar2023AEM, Kan2013APL} This protects the junction by making it more insulating, as indicated by an enlarged internal field~(Fig.~\ref{fig:17}e). In contrast, a positive voltage (forward bias) promotes the charge detrapping that leads to the LRS; however, if a sufficiently large voltage is applied, the large current generated will lead to a soft electroforming process through Joule heating.~\cite{Yang2014JAP, Baeumer2016N} Thus, the large positive voltage regime changes the RS mechanism to CF-type. A I$_\mathrm{CC}$ is typically used to protect M/Nb:STO devices in the large positive bias regime from degradation and CF formation.~\cite{Li2018PSSA, Baeumer2016N} Overall, voltage bias not only controls the Schottky barrier, but also governs the underlying switching mechanism.

\section{Neuromorphic Applications}

Neuromorphic computing requires devices capable of mimicking the structure and function of biological neurons and synaptic connections. The computational value of the M/Nb:STO platform is not merely RS, but that the RS is governed by a tunable interface. Trapped electronic charge and interfacial chemical species modulate a rectifying Schottky barrier whose $C_\mathrm{i}$ and $C_\mathrm{d}$ can be set by fabrication and conditioning.~\cite{Buzio2012APL, Bian2025FML, Mikheev2014NatureCommunications, Kunwar2023AEM} This co-locates three levers in a single element: a built-in selector (intrinsic rectification), a slow chemical state (OH$^-$ or O$^{-2}$) that acts as context, and a $T$-dependent permittivity $\epsilon_s$($T$) that retunes depletion/tunneling without reprogramming the stored conductance when reads are brief and low-duty, the latter is consistent with STO’s paraelectric dielectric response.~\cite{Muller1979PRB, Rowley2013NP}. The result is a single material stack that can be operated as a weight in a crossbar, a context-gated synapse, or a volatile dynamical node, purely by biasing and interface preparation. Across these use cases the through-line is explicit, computation follows directly from the interface physics. Here we discuss the potential application of Nb:STO with an eye towards unique functionality gain from the tunable interface.


\subsection{Interface Tunability}

The interface-type switching in Nb:STO is advantageous for its tunability between non-volatile and volatile storage capabilities. This can be tuned via electronic protocols (e.g., voltage polarity, amplitude, pulse width, \textit{etc}.) and interface quality (e.g., electrode material/deposition, surface treatment, Nb-concentration, \textit{etc}.) for a specific application.~\cite{Quinonez2025APLED, Kunwar2023AIM} For example, the retention of an Au/Nb:STO device measured at $\pm 0.4~V$ decays with $\beta = 0.343$ for negative polarity and $\beta = 0.016$ for positive polarity, and a larger positive bias leads to better LRS retention than relatively smaller positive bias reads.~\cite{Kunwar2023AEM} Additionally, the trapping site population, which is tuned by interface chemistry, will impact the magnitude of RS but also the timescales of SET/RESET operations and retention dynamics.~\cite{Li2018PSSA, Lee2014APLM, Mikheev2015SR, Quinonez2025APLED, Zhang2010APL} Therefore, the sensitivity of M/Nb:STO junctions directly leads to tunable memristor performance. 

The primary method for tuning the conductance of M/Nb:STO devices in neuromorphic applications is through voltage pulses, defined by the polarity, magnitude, duration, and duty. By applying successive pulses, potentiation (i.e., increasing conductance) can be achieved with positive voltage, while depression (i.e., decreasing conductance) with negative voltage. In terms of conductance modulation, it was found that a writing pulse with increasing magnitude and duration lead to enhanced potentiation in Au/, Ni/, and Co/Nb:STO.~\cite{Kunwar2023AIM,Quinonez2025APLED, Tiotto2021FN} The potentiation response becomes increasingly linear when the write-pulse amplitude is progressively increased,~\cite{Kunwar2023AIM} whereas constant-amplitude write pulses typically produce power-law conductance evolution.~\cite{Quinonez2025APLED, Tiotto2021FN, Kunwar2023AIM} Controlling the linearity of potentiation and depression remains a major challenge for artificial synapses to match the performance of numerical neural networks.~\cite{Gokmen2016FN, Chang2017JESTCS} Thus, optimizing the M/Nb:STO fabrication and pulse protocol via co-design for linear conductance updates is a frontier research direction. 

The time domain also provide a method for controlling synaptic weights. An increasing time between pulses lead to suppressed paired-pulse facilitation, where the second pulse has less impact than the first pulse.\cite{Kunwar2023AIM}
STDP has also been shown in M/Nb:STO, where the time difference between the pre-synaptic and post-synaptic pulses controls the sign and magnitude of the conductance change.\cite{Kunwar2023AIM}
Moving on to retention, the resistance state following a write pulse initially decays quickly, corresponding to short-term plasticity, and then saturates to a stable value, corresponding to a non-volatile effect.~\cite{Quinonez2025APLED} Quinonez \textit{et al}. used a numerical OV migration model to show the volatile effect arises from OV diffusion from the depletion layer to bulk Nb:STO, while the non-volatile effect is induced by a permanent redistribution of OVs near the interface.~\cite{Quinonez2025APLED} Further, Kunwar \textit{et al}. demonstrated that the retention increases for an increasing number of successive pulses.~\cite{Kunwar2023AIM} This can be extended where short forward-bias pulse trains yield short-term potentiation/depression (STP/STD), whereas longer, low-bias exposures consolidates long-term retention (LTP/LTD). The temporal structure and history of electrical stimuli therefore provide an additional degree of control over the synaptic state.

The chemical degree of freedom of M/Nb:STO Schottky contacts is not a peripheral sensing artifact but a controllable channel for in-sensor learning: RS and Schottky barrier modulation originate in an interfacial layer whose properties (e.g., $n$, $C_\mathrm{i}$, $\delta$) are set by local chemistry, and they are reversibly tuned by ambient oxygen/water, including explicit identification of protons from moisture as the critical species in M/Nb:STO interface-type memristors.~\cite{Kunwar2023AEM, Quinonez2025APLED, Buzio2012APL, Bian2025FML} Under controlled ambient conditions, longer pulses drive proton/oxygen redistribution that enhance the non-volatile retention characteristics.~\cite{Quinonez2025APLED, Goossens2018JAP} The STP and LTP separation and its oxygen/proton dependence mirror results in solid-state protonic synapses and oxide devices where the proton chemical potential sets learning kinetics.~\cite{Yao2020NC, Buzio2012APL, Quinonez2025APLED} Collectively, these studies support treating ambient-controlled interfacial chemistry as a first-class knob for context-gated synapses in Nb:STO-based systems. As these knobs are physically distinct, their time constants can be tuned independently. 

Temperature can be considered another control knob, which can be treated as a calibration parameter. The permittivity $\epsilon_s$($T$) of STO enters the electrostatics of a Schottky contact through the $W_\mathrm{d}$ and image-force barrier lowering, so modest, rapid $\Delta T$ produces smooth shifts in the small-signal current and dI/dV at a fixed read bias.~\cite{Coak2018SR} In contrast, trap/proton occupancy evolves on much longer timescales at low (high) negative (positive) read amplitude and duty, so brief reads change "gain" without moving the stored state. State drift induced by repeated low-amplitude read becomes measurable only under sustained or higher-bias stress, consistent with read-disturb behavior reported in RRAM reliability studies.\cite{Wang2022SSE,Zhao2020APR} This is consistent with the strong, smooth dielectric variation of STO at low $T$, thus motivating characterizing the read transfer as a function of $T$ and explicitly delimiting a “no-reprogramming window” (read voltage, duty cycle, $\Delta T$). The non-linear field dependence of $\epsilon_s (T,E)$ and voltage partitioning in the MIS model give the appearance of reducing $\epsilon_s$ at low-$T$,~\cite{Susaki2007PRB, Goossens2018JAP,Kim2020APL} thus the no-programming window must be carefully tuned.~\cite{Shimizu1999JAP, Rana2013PRB} It follows, one could implement $T$-coded features: the same physical weight responds differently to the same stimulus as $T$ varies, aiding sensor function. Thus enabling applications wherein $T$-coded features relate to device or circuit performance, or within cryogenic interfaces.~\cite{Islam2023JAP,Chen2023NR, Torres2023AM}

\subsection{Non-volatile Implementation}

\begin{figure}[t]
    \centering
    \includegraphics[width=\linewidth]{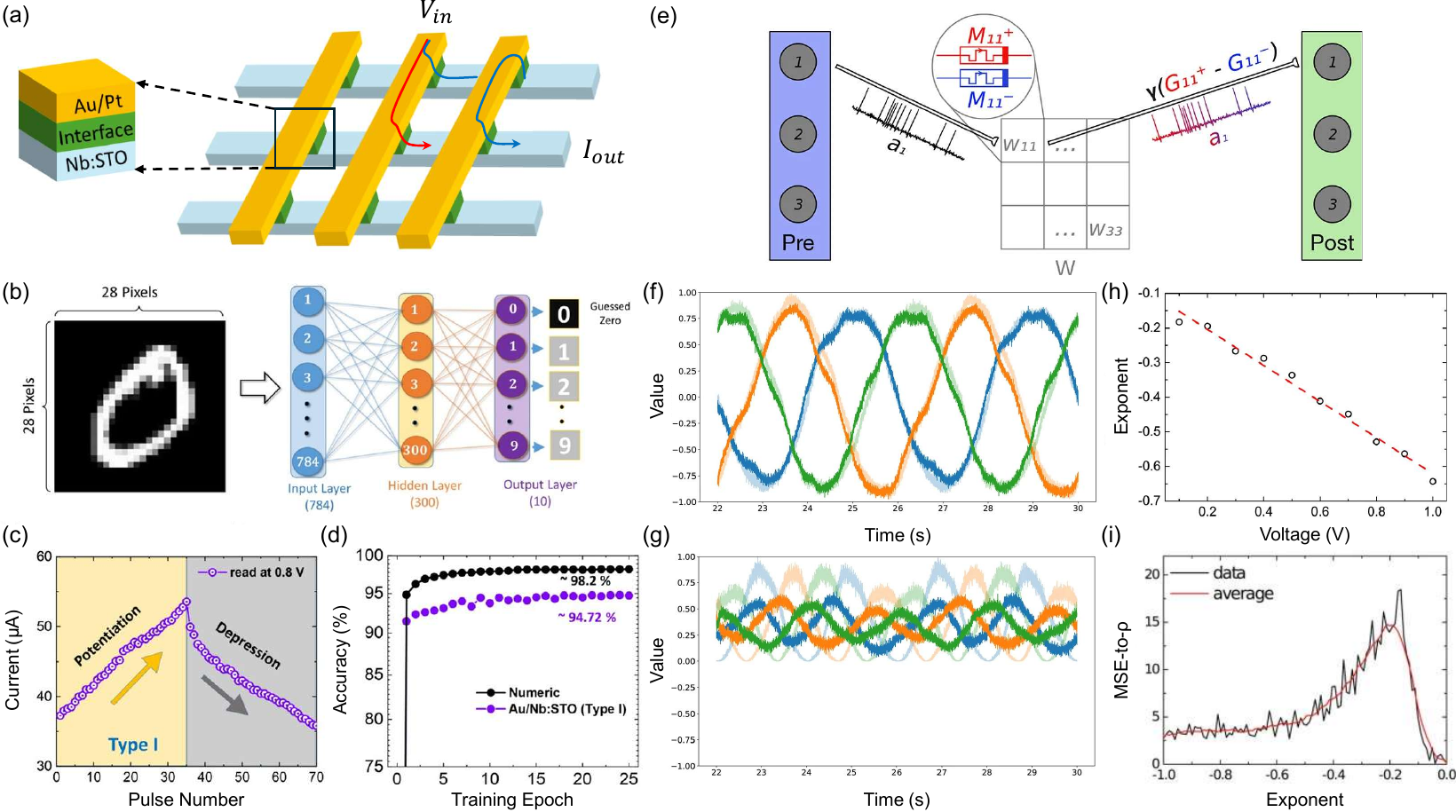}
    \caption{(a) Diagram of an (Au, Pt)/Nb:STO crossbar array with interface-type switching. The normal current path for $V_\mathrm{in} \rightarrow I_\mathrm{out}$ is shown as the red arrow, while the sneak path is depicted with the blue arrow. (b) Schematic of simulated neural network based on Au/Nb:STO electrical characterization.  (c) Training protocol output current demonstrating potentiation and depression in Au/Nb:STO. (d) Comparison between simulated Nb:STO neural network and numerical categorization of handwritten numbers. (e) Schematic of simulated spiking neural network, involving leaky integrate and fire (LIF) pre-synaptic encoder neurons, a weight matrix composed of simulated Ni/Nb:STO memristors,
    and post-synaptic decoder neurons. For the weight matrix, $i$ indicates the pre-synaptic neuron and $j$ the post-synaptic decoder neuron. The decoded pre-synaptic (faded color) and post-synaptic (solid color) signals for learning (f) $f(x)=x$ and (g) $f(x)=x^2$, based on a 3$d$ sinusoidal input signal.(h) Power-law decay exponents for varying pulsing voltages with a duration of 1 second. (i) Mean squared error (MSE) to Spearman correlation coefficient ($\rho$) ratio (i.e., $\rho / \mathrm{MSE}$) for different exponents used to update synaptic weights.
    Figures (b-d) taken from Ref.~\cite{Kunwar2023AIM}, and (e-h) from Ref.~\cite{Tiotto2021FN}. }
    \label{fig:19}
\end{figure}

Crossbars are the natural starting point because they are the standard layout for ReRAM and for analog vector–matrix multiplication: rows apply voltages, columns collect currents, and Kirchhoff’s laws perform the multiply-accumulate.~\cite{Ielmini2018NE} The perennial problem is leakage current from unaddressed cells, as there are natural sneak paths that span neighboring rows and columns (see Fig.~\ref{fig:19}a). Here the rectifying M/Nb:STO contact functions as a selector by design, aligning with the broader class of self-rectifying memristors shown to suppress sneak currents and even enable selector-less passive arrays at scale.~\cite{Jeon2024NC,Jeon2021NC,Shi2020NA,Chen2010APL,Chen2022FEM} 
The M/Nb:STO junction is intrinsically nonlinear,  its I-V has a pronounced non-linearity, i.e., a "knee" at a voltage $V_\mathrm{knee}$.  In practice we operate at read biases well below V$_\mathrm{knee}$ and evaluate array-level accuracy under a standard V/2 scheme.~\cite{Li2021AIS, Chen2024SR} Lastly, interface-type switching possesses more reproducible and tunable switching relative to CF-type, due to interface and defect engineering to optimize the trapping/detrapping mechanism.~\cite{Shooshtari2025AIS}

The interface-controlled RS in M/Nb:STO enables analog and gradual conductance updates, which are key requirements for implementing in LTP and LTD. In biological synapses, the LTP and LTD correspond to persistent strengthening and weakening of synaptic weights following repeated stimulation, which the Au/NbSTO device can emulate with voltage pulses. In the Au/Nb:STO device, consecutive positive voltage pulses induce gradual conductance enhancement (i.e. LTP), while negative pulses produce progressive conductance reduction (i.e. LTD). Since the real device response is generally non-linear (non-ideal), the LTP/LTD response (linearity and symmetry) also deviates from ideal devices. An optimized programming pulse scheme can bring linearity and symmetry closer to the ideal device~(Fig.~\ref{fig:19}c). This is advantageous for neural network training and inference. Kunwar \textit{et al}. evaluated the system-level performance of Au/Nb:STO by simulating a crossbar array using CrossSim~\cite{Plimpton2016}~(Fig.~\ref{fig:19}b).~\cite{Kunwar2023AIM} The simulated artificial neural network achieved above 94~\% accuracy for MNIST image classifications~(Fig.~\ref{fig:19}d). For comparison, an ideal linear numeric model achieved 98~\% accuracy, indicating that the slight nonlinearity and asymmetry of practical devices introduce modest performance degradation. Nonetheless, the achieved accuracy compares favorably with filamentary ReRAM synapses.~\cite{Jacobs2018IEEETNS,Bennett2019IEEEIRPS,Ma2022ACSAMI} Importantly, the tight distribution of conductance change ($\Delta$G vs. G) during LTP/LTD measured for cycle-to-cycle and device-to-device variation leads to stable weight updates and improved learning.  

To evaluate the suitability of M/Nb:STO memristors for functional approximation tasks, Tiotto \textit{et al.} used experimental potentiation curves of Ni/Nb:STO memristors to simulate non-volatile synaptic weights in a spiking neural network (SNN).~\cite{Tiotto2021FN} As illustrated in Fig.~\ref{fig:19}e, pre-synaptic neurons encode an input signal into spike trains, while synaptic weights are implemented using differential memristor pairs with an effective weight proportional to $G^+ - G^-$. The post-synaptic layer then decodes the resulting spike activity to reconstruct the target function.~\cite{Tiotto2021FN} Using a network of 100 pre- and post-synaptic neurons, Tiotto \textit{et al.} demonstrated successful approximation of both linear and nonlinear functions (Fig.~\ref{fig:19}f,g).~\cite{Tiotto2021FN} The linear function was reproduced more accurately, indicating that the network more readily learned simple input-output relationships. Learning performance was strongly influenced by the nonlinear conductance-update behavior of the memristors. The conductance update exponent varied approximately linearly with pulse voltage (Fig.~\ref{fig:19}h), while optimal functional-approximation performance was achieved for exponents near $-0.17$ to $-0.16$, as indicated by the maximum MSE-to-$\rho$ value (Fig.~\ref{fig:19}i).~\cite{Tiotto2021FN} These results demonstrate that the non-volatile conductance states of M/Nb:STO memristors can serve as analog synaptic weights in SNN hardware, enabling learning and functional approximation tasks relevant to neuromorphic computing applications.

Inter-device variability is the limiting factor; we counter it with write–verify and hardware-aware training.  We note that 1S1R (i.e., one-selector-one-resistor) deployment on Nb:STO has been limited in large arrays due to the strong interfacial and ambient sensitivity of Schottky-controlled RS and the availability of mature oxide selectors that meet V/2 crossbar requirements with higher reproducibility. Given the interfacial control, there is potential for highly controllable 1S1R crossbar arrays with mature Nb:STO stacks.

\section{Conclusion and Opinion}

The M/Nb:STO junctions have been widely studied in the past several decades owing to forming-free, analog RS characteristics and large on/off ratio.  The intrinsic M/Nb:STO junction, while exhibiting current rectification, is essentially free of RS. Instead, the emergence of RS is closely associated with the formation of an extrinsic interfacial layer between the metal and Nb:STO,~\cite{Mikheev2014NatureCommunications} effectively transforming the junction from a metal–semiconductor (MS) structure into a metal–insulator–semiconductor (MIS) structure.~\cite{Hasegawa1991JAP,Yoshida1991PhysicaB,Shimizu1999JAP,Yamamoto1998JJAP,Susaki2007PRB} This review has summarized the dominant interpretations of RS being charge trapping/detrapping, tunneling, and OV-based mechanisms across these MIS junctions. Because the RS is related to the extrinsic interfacial layer formed between the metal and Nb:STO, the switching properties of this junction then largely depend on the metallic deposition, Nb:STO surface quality and chemistry, and aging conditions.  Therefore, it is not surprising that a variety of I-V hysteresis loops observed and different mechanisms proposed. When we anchor RS with the extrinsic interfacial layer with spatial inhomogeneity, these mechanisms can be reconciled into a unified physical picture. 

The SBH modulation mechanism captures the physical picture how charge trapping/detrapping at charged defects at or near the extrinsic interfacial layer results in forming-free RS in M/Nb:STO.~\cite{Brillson2011JAP,Dharanya2022JNP,Bourim2013CAP,Buzio2012APL,Bian2025FML,Bourim2014JSSST,Chen2011APL,Park2014APL,Lee2014APLM,Quinonez2025APLED,Shen2013APA,Zhong2013CAP,Chen2010APL,Mikheev2014NatureCommunications,Buzio2024JoPDAP,Fan2017JoMCC,Park2008JAP,Li2018PSSA,Goossens2018JAP,Kunwar2023AEM,Yin2015PCCP,Kan2013APL,Li2019PSSA} The modulation of the SBH and $W_\mathrm{d}$ naturally leads to a transition in transport, where the HRS (LRS) is dominated by thermionic emission (tunneling).~\cite{Fan2017JoMCC} Both the barrier modulation and transport is strongly sensitive to the extrinsic interfacial layer formation, which is typically enhanced with low energy metal deposition (e.g., room-$T$ evaporation) and moderately doped Nb:STO substrates.~\cite{Mikheev2014NatureCommunications,Li2019PSSA, Chen2011APL}  As the charge trapping/detrapping relies on an extrinsic interfacial layer, poor metallic contact and various surface defects are what facilitate RS. More recently, aging studies have shown that moisture (i.e., protonic defects) is a key species for charge trapping/detrapping in M/Nb:STO.~\cite{Kunwar2023AEM} It is important to note that, within the SBH-modulation mechanism, the SET (RESET) voltage must be sufficiently large to induce effective electron detrapping (trapping), while remaining below the threshold for electroforming. For example, forming-free RS is typically maintained by limiting the junction current to~10 mA, while relatively large voltage excursions may still be required to fully access the LRS (i.e., 3~$V$) and HRS (i.e., -6~$V$).~\cite{Kunwar2023AEM, Li2018PSSA}  Due to this bipolar RS, the LRS retention is enhanced when measuring with a positive DC bias by promoting the detrapping of charge.~\cite{Tian2011APA,Bourim2014JSSST,Zhang2009APL, Goossens2018JAP, Mikheev2014NatureCommunications,Ni2007APL, Fan2017JoMCC,Kunwar2023AEM} Forming-free RS is the dominantly reported device behavior in M/Nb:STO, and is characterized by an extrinsic RS mechanism that is sensitive to fabrication, storage, and measurement conditions. 

SBH inhomogeneity (Fig.~\ref{fig:22}b) provides a natural explanation for the apparent discrepancy between the SBH-modulation mechanism and the minimal overall SBH changes between the HRS and LRS observed by capacitance and photoelectric measurements.~\cite{Wang2013APL,Shang2008APL, Lee2011APL} If RS occurs only within a small fraction of the nominal electrode area, substantial local modulation of the SBH can produce only a minor change in the spatially averaged barrier height. Thus, the absence of a measurable global SBH change in photoelectric measurements does not necessarily rule out significant local SBH modulation as the origin of RS.~\cite{Shang2008APL, Lee2011APL} 

Therefore, a more realistic physical picture of forming-free RS in M/Nb:STO can be proposed. The formation of an extrinsic interfacial layer between the metal and Nb:STO is central to the observed RS, with modulation of the SBH and/or $W_\mathrm{d}$, through charge trapping/detrapping playing a dominant role in controlling charge transport. Electron trapping increases the effective SBH and/or $W_\mathrm{d}$, giving rise to the HRS, whereas detrapping lowers the interfacial barrier, enhances tunneling, and drives the device into the LRS. Importantly, in contrast to the conventional picture in which forming-free interface-type RS occurs uniformly across the entire electrode area, experimental evidence suggests that only a fraction of the nominal electrode area is electrically active during switching. The fraction of this active area remains difficult to quantify and likely depends strongly on interface quality and fabrication conditions.

An important open question is the extent to which this physical picture can be generalized from M/Nb:STO to M/oxide/Nb:STO heterostructures. The striking similarities in their I–V characteristics and hysteresis rotation sequences suggest that the interfacial mechanisms established for M/Nb:STO may also provide a useful framework for understanding M/oxide/Nb:STO devices. However, the additional oxide layer introduces greater complexity: either the top M/oxide interface or the bottom oxide/Nb:STO interface has been proposed as the dominant switching region,~\cite{Lee2014AM,Yun2021JMCC} while both OV migration and defect-mediated charge trapping/detrapping (including trapping associated with OVs) have been invoked as possible switching mechanisms.~\cite{Xia2020AIPA,Li2021PSSA,Su2024AMI,Kim2026CAP} Systematic comparisons between M/Nb:STO and M/oxide/Nb:STO are therefore needed to distinguish these contributions and determine which aspects of the proposed framework are universal. In particular, comparing their I–V hysteresis and rotation sequences, retention, endurance, and dependence on measurement protocols could provide important mechanistic insights. Equally important is determining how sensitively RS in M/oxide/Nb:STO depends on the chemistry and quality of the top M/oxide and bottom oxide/Nb:STO interfaces. Such interface sensitivity is relatively established in M/Nb:STO junctions but remains largely unexplored in M/oxide/Nb:STO heterostructures. It is reported that water moisture is strongly connected to charge trapping process in M/Nb:STO devices, while such an effect in M/oxide/Nb:STO has largely unexplored. Very recent work has shown moisture also plays an important role in diffusive memristors based on ionic effects.~\cite{Kim2026NCE}

The post-forming filament-type RS is a distinct regime that appears when M/Nb:STO junctions are biased above approximately 10 mA.~\cite{Yang2014JAP, Baeumer2016N}  Due to the filament formation, the post-forming M/Nb:STO junction possesses a reduced effective area and localized VCM-type switching. To explain the CC-C rotation sequence, the RS is achieved by a combination of OV-drift and interfacial oxygen exchange.\cite{Cooper2017AM, Yang2014JAP,Baeumer2016N} A positive bias achieved the LRS by having oxygen ions migrate to the M/Nb:STO interface and leave the lattice, reducing the Nb:STO filament.~\cite{Baeumer2016N} For a negative bias, the filament is re-oxidized, leading to the HRS.~\cite{Baeumer2016N} The VCM-type filaments represent another facet of M/Nb:STO device, which is achieved through high bias forming, and is less explored compared to the forming-free RS. The filamentary-type VCM in M/Nb:STO discussed here and many other M/oxide/Nb:STO devices are the same.

The M/Nb:STO junctions host a forming-free interface-type RS capable of analog conductance updates, representing a foundational system to explore and optimize synaptic functionality. While near linear potentiation and depression curves have been shown using a tailored pulse protocol,~\cite{Kunwar2023AIM} future work should explore interface engineering to optimize both the linearity and symmetry. A systematic exploration of fabrication, environmental, and measurement conditions may further optimize the retention, endurance, and LTP/LTD required for practical artificial synapses. The lessons learned from such studies may be applied to more general interface-type memristors, and even hybrid interface- and filament-type memristors.~\cite{Dou2023AEM,Bakhit2026SA} 
Further, such interface engineering will enable functional Nb:STO heterostructures in crossbar array architectures. The similar RS characteristics observed in M/oxide/Nb:STO and M/Nb:STO junctions suggest that the mature M/Nb:STO system can provide a baseline for understanding how the M/oxide and oxide/Nb:STO interfaces, together with the functional oxide layer, modify device behavior.~\cite{Fan2017JoMCC,Zhao2019RSOS,Wang2024JPCL,Xie2019JAC} These insights will guide the integration of mature Nb:STO stacks with ferroelectrics, high-$\kappa$ dielectrics, multiferroics, optoelectronics, magnetoelectrics, and other functional perovskite oxides.

\section{Acknowledgments}
The work at Los Alamos National Laboratory was supported by the NNSA's Laboratory Directed Research and Development Program, and was performed, in part, at the CINT, an Office of Science User Facility operated for the U.S. Department of Energy, Office of Science. Los Alamos National Laboratory, an affirmative action equal opportunity employer, is managed by Triad National Security, LLC for the U.S. Department of Energy's NNSA, under contract 89233218CNA000001. This work is also partially supported by the U.S. Department of Energy, Office of Science, Basic Energy Sciences, as part of the Microelectronics Energy Efficiency Research Center for Advanced Technologies (MEERCAT), a Microelectronics Science Research Center (MSRC). Authors also acknowledge the support from DOE ASCR Express.

\bibliography{bib}

\end{document}